\documentclass{statsoc}

\usepackage[a4paper]{geometry}
\usepackage{graphicx}
\usepackage[textwidth=8em,textsize=small]{todonotes}
\usepackage{amsmath}
\usepackage{natbib}
\usepackage{verbatim}
\usepackage{float}

\title[Posterior Summaries of Grocery Retail Topic Models]{Posterior Summaries of Grocery Retail Topic Models:
 \\Evaluation, Interpretability and Credibility}
\author{Mariflor Vega}\address{University College London, London, UK} \email{mariflor.vega.15@ucl.ac.uk}

\author{Ioanna Manolopoulou} \address{University College London, London, UK}

\author{Jason O'Sullivan} \address{dunnhumby Ltd, London, UK}

\author{Rosie Prior} \address{dunnhumby Ltd, London, UK}

\author[Mariflor Vega {\it et al.}]{Mirco Musolesi} \address{University College London, London, UK}

\usepackage{tikz}
\usetikzlibrary{fit,positioning}
\graphicspath{ {./images/} }
\usepackage{url}
\usepackage{subcaption}
\usepackage{tabulary}
\usepackage{multirow}
\usepackage{bbm}

\begin{document}
%%%%%%%%%%%%%%%%

\begin{abstract}

Understanding the shopping motivations behind market baskets has significant commercial value for the grocery retail industry. The analysis of shopping transactions demands techniques that can cope with the volume and dimensionality of grocery transactional data while delivering interpretable outcomes. Latent Dirichlet Allocation (LDA) provides a natural framework to process grocery transactions and to discover a broad representation of customers' shopping motivations. However, summarizing the posterior distribution of the LDA model is challenging. LDA is, in essence, a mixture model and thus, even after addressing the label-switching problem, averaging topic distributions across posterior draws may fuse topics that illustrate distinct shopping motivations. On the other hand, individual LDA draws cannot capture the uncertainty around topics, as topics may (dis)appear in different posterior samples. In this paper, we introduce a clustering methodology that post-processes posterior LDA draws to summarise the entire posterior distribution and identify semantic modes represented as recurrent topics. Instead of resolving the label-switching problem by re-labelling components from each posterior draw to minimize an overall loss function, our approach identifies clusters of component parameters across iterations. This allows us to potentially identify multiple modes (termed topical themes) in the posterior distribution, as well as provide associated measures of uncertainty for each cluster. We also establish a more holistic definition for model evaluation, which assesses topic models based not only on their predictive likelihood but also on qualitative aspects such as coherence and distinctiveness of individual topics and credibility of the set of identified topics. Using the outcomes of a tailored survey, we set thresholds that aid the interpretation of the qualitative aspects in the domain of grocery retail data. We demonstrate that the selection of recurrent topics through our clustering methodology not only improves predictive likelihood but also outperforms qualitative aspects such as interpretability and credibility. We illustrate our methods on an example from a large British supermarket chain.

\end{abstract}

\maketitle
% Samples of sectioning (and labeling) in MNSC NOTE: (1) \section and \subsection do NOT end with a period (2) \subsubsection and lower need end punctuation (3) capitalization is as shown (title style).
%
% \section{Introduction.}\label{intro} %%1.
% \subsection{Duality and the Classical EOQ Problem.}\label{class-EOQ} %% 1.1.
% \subsection{Outline.}\label{outline1} %% 1.2.
% \subsubsection{Cyclic Schedules for the General Deterministic SMDP.}
% \label{cyclic-schedules} %% 1.2.1
% \section{Problem Description.}\label{problemdescription} %% 2.

% Text of your paper here

\section{Introduction}
In the grocery retail industry, millions of transactions are generated every day out of thousands of products available to customers. Understanding combinations of products that are purchased together can help us to characterize customer needs and profiles. For example, shopping needs may reflect foods consumed at breakfast, ingredients for roast dinner, or products for a barbecue; customer profiles may correspond to specific types of products such as organic products or luxury branded items to fulfil shopping needs. Modelling these combinations is typically treated as an unsupervised learning task, essentially amounting to a high-dimensional clustering problem, where clusters represent weighted product combinations. Subject-matter expertise can then provide an explicit assessment of the quality of an individual cluster or set of clusters in terms of grocery retail interpretation. The motivation of this paper is two-fold: first, to provide a framework in which we can combine existing, well-studied Bayesian statistical topic models with subject-matter evaluations of model quality measured through tailored user surveys in grocery retail; second, to introduce a posterior distribution summary which can simultaneously capture the posterior mean and quantify posterior uncertainty of a topic model.

Topic modelling (TM) is a natural, scalable statistical framework that can process millions of combinations of products within transactions while maintaining the explanatory power to discover, analyze, and understand customer behaviours. Latent Dirichlet Allocation (LDA) \citep{blei2003latent} is one of the most popular topic modelling techniques. In the context of retail data, LDA represents transactions as topical mixtures, where each topic is a multinomial distribution over the product assortment. LDA interprets each transaction as an unordered \textit{bag of products}, which is natural in the grocery retail domain as products are registered without an inherent order. In this formulation, a transaction is a probabilistic realization of a mixture model over shopping `goals' (for example, ingredients for a roast and baby food) \citep{hornsby2019conceptual}. The analysis of grocery transactions through the application of topic modelling can provide insights into shopping patterns at a much higher resolution, which may enhance cross-selling campaigns, product assortments and layouts, and may aid the analysis of social and cultural habits of grocery purchasing.

Topic model evaluation is typically based on model fit metrics such as held-out-likelihood or perplexity \citep{wallach2009evaluation,buntine2009estimating}, which assess the generalization capability of the model by computing the model likelihood on unseen data. However, LDA likelihood does not capture qualitative aspects such as semantic \textit{topic coherence} \cite{newman2010automatic}, and hence these metrics may lead to topic models with less semantically meaningful topics according to human annotators \citep{chang2009reading}. Topic coherence was introduced to measure the difficulty of associating an individual topic from an LDA posterior draw to a single semantic concept, and consequently, evaluating topic models by their interpretability. Topic coherence is typically quantified by co-occurrence metrics such as Pointwise Mutual Information (PMI) and Normalized Pointwise Mutual Information (NPMI) \citep{bouma2009normalized}, which have been shown to correlate with human annotators in \cite{newman2010automatic, lau2014machine}. Meaningful implementation of these metrics requires knowledge about their relevance to human perception of topics in the domain of interest -- in our case, mapping measurements of PMI and NPMI to the quality of the corresponding shopping goals represented by the topics.

Topic coherence or held-out metrics do not capture similarity among topics within a sample or posterior variability across samples. Topics within the same posterior sample may contain product combinations that could be associated with the same semantic concept \citep{boyd2014care}, but whose variations prohibit the likelihood model from merging them. Since topic models are meant to explain a corpus, having semantically similar topics within the same posterior sample is a suboptimal outcome \citep{boyd2014care}. Topics may also exhibit significant variations across a set of posterior samples,  \citep{steyvers2007probabilistic,rosen2010learning,chuang2015topiccheck}; a topic associated with a particular semantic concept (in our case, a shopping goal) may appear and disappear across multiple posterior samples, depending on its posterior uncertainty. In response, we establish a more holistic definition for model evaluation, which assesses topic models based not only on their likelihood but also on qualitative aspects such as \textit{topic coherence}, \textit{topic distinctiveness} and \textit{topic credibility}. Topic distinctiveness measures semantic dissimilarity among topics. Topic credibility quantifies semantic similarity among posterior samples. Thus, topics of high quality are not only coherent, but also distinctive within, and recurrent among, posterior draws. 

Topics of low uncertainty may appear consistently across posterior samples and reflect `reliable' topics, whereas topics that do not appear consistently among LDA posterior draws may correspond to idiosyncratic product combinations. However, explicit quantification of this variability is challenging because there is currently no unique measure of similarity between topics or topic models, or universal thresholds that determine when a pair of topics are (dis)similar. At the same time, computing component-wise posterior means is non-trivial because topic models are in essence mixture models and thus subject to the label switching problem  \citep{stephens2000dealing,jasra2005markov,sperrin2010probabilistic,hastie2015sampling}; even after resolving component labels, computing posterior summaries of a multi-modal posterior will frequently average across component draws which represent different semantic concepts.

In this work, we develop a post-processing methodology that aggregates multiple posterior samples of LDA to capture a single summary of semantic modes, presented as a set of recurrent topics that also contains a measure of uncertainty. Rather than assigning one-to-one matches of topics across posterior draws as in standard label-switching approaches, we use hierarchical clustering to group topics from different posterior samples according to a distance criterion for topic distributions. Here, we use cosine distance among distributional measures as it correlates with human judgment on topic similarity \citep{aletras2014measuring}. A \textit{clustered topic} is then defined as the average topic distribution from different LDA posterior samples that exhibit the same theme, and its posterior credibility is measured as the topic recurrence, i.e., the number of topics within the cluster (the number of posterior samples exhibiting the same topic). Guided by the domain of interest, users can tune the distance criterion and set thresholds of minimum recurrence to select clustered topics of low uncertainty. The resulting clustered LDA model is used to lead interpretations on customers' needs instead of using a single LDA posterior draw or a posterior mean (thus circumventing the need to address the label-switching problem across different samplers).

We present a user study in which experts in grocery retail analytics assessed topics for their interpretability and similarity. We use this study to relate our measures of topic quality to users' intuitive perception of these concepts. We interpret LDA topics in the application to grocery retail data and show that some LDA topics (within a posterior sample) may not be the most coherent, distinctive, and credible. Moreover, we demonstrate that the selection of recurrent topics through our clustering methodology provides subsets of clustered topics with better model likelihood, greater credibility and improved interpretability.

This paper is organized as follows: we discuss related work in Section \ref{sec:litreview}. LDA is described in Section \ref{sec:LDA}. Section \ref{sec:evaluation} presents the definitions of model generalization, topic coherence, topic distinctiveness, and topic credibility. Section \ref{sec:LDAclustering} introduces our proposed methodology for clustering and selecting recurrent topics. Sections \ref{sec:groceryuserstudy}, \ref{sec:groceryclustering}, and \ref{sec:grocerydiscussion} show the application of grocery retail data from a major retailer in the UK. More specifically, Section \ref{sec:groceryuserstudy} discusses thresholds for interpretability and similarity obtained from a user study with experts in grocery retail analytics and exhibits the pitfalls of LDA topics. Section \ref{sec:groceryclustering} demonstrates the advantages of selecting clustered topics of high posterior recurrence. Section \ref{sec:grocerydiscussion} displays identified grocery topics and indicates commercial implications in the grocery retail sector. Finally, we summarise our findings in Section \ref{sec:conclusion}.

\section{Related work\label{sec:litreview}}

Topic modelling, in particular LDA, has already been used to identify latent shopping motivations in retail data. For instance, \cite{christidis2010exploring} applied LDA to grocery transactions from a major European supermarket to identify latent topics of product categories, intending to support an item recommendation system. In this study, 102 thousand unique products were aggregated into 473 synthetic categories with no distinction between brands or package sizes. \cite{hruschka2014linking} sketched the core of a recommender system to illustrate the managerial relevance of estimated topics, which were obtained from training LDA and the correlated topic model on market baskets from a medium-sized German supermarket. The study only accounted for the 60 product categories with the highest univariate purchase frequencies. \cite{jacobs2016model} applied topic models to market baskets from a medium-sized online retailer in the Netherlands to identify latent motivations and to predict product purchasing in large assortments. Again, the authors aggregated products to a category-brand level, i.e., different fragrances/flavours of the same product and brand are aggregated into one category, reducing more than 3 thousand unique products to 394 categories. \cite{hruschka2016hidden, hruschka2019comparing} compared topic models and other unsupervised probabilistic machine learning methods on point-of-sale transactions from a typical local grocery store in Austria, analyzing 169 product categories. The aforementioned works analyzed collections of product categories and not the full product resolution, thereby reducing the dimensionality of the problem. \cite{ hornsby2019conceptual} provided a direct application of a 25-topic LDA model on transactional data from a major British retailer to identify shopping goals.

Beyond LDA, other approaches have been applied to market baskets. For instance, \cite{schroder2017using} applied Multidimensional Item Response Theory (MIRT) models on a limited dataset with 31 product categories collected by a house panel from a single supermarket in the US, and found that MIRT models outperformed LDA according to the Akaike Information Criterion and its corrected form. MIRT may be an option to analyze small datasets of discrete grouped data. \cite{hruschka2016hidden, hruschka2019comparing} also compared topic models such as LDA and correlated topic model (CTM) to alternative methods such as binary factor analysis, restricted Boltzmann machine (RBF), and deep belief net (DBN). It was shown that the alternative methods outperform topic models in model generalization. However, the number of topics was restricted to a range from 2 to 6, while networks of much larger architectures were explored. Moreover, the DBN and RBF outcomes are far less interpretable than LDA topics.
\cite{ruiz2020shopper} introduced `SHOPPER', a sequential probabilistic model, that captures interaction among items and answers counterfactual queries about changes in prices. \cite{chen2020studying} introduced `Product2Vec', a method based on the representation learning algorithm Word2Vec, to study product-level competition, when the number of products is large and produce more accurate demand forecasts and price elasticities estimations. \cite{jacobs2020understanding} combined the correlated topic model with the vector autoregression to account for product, customer, and time dimensions present in purchase history data. 

Within LDA, various methods have been proposed to improve topic coherence. For example, \cite{wallach2009rethinking} used asymmetric priors over document distributions to capture highly frequent terms in few topics; \cite{newman2011improving} introduced two regularization methods, and \cite{mimno2011optimizing} generalized the P\`olya urn model aiming to reduce the number of low-quality topics. In this paper, we do not try to improve LDA to render more coherent topics, but we will show that our proposed methodology retrieves groups of clustered topics with higher coherence. 

Hierarchical clustering has been used previously to interactively align topics \citep{chuang2015topiccheck} and to aggregate topic models \citep{blair2016increasing}. The former work assumes that topics align with up to one topic from a different posterior sample. The latter work merges topics from posterior samples with small and large numbers of topics aiming to improve topic coherence. However, these works do not assess other aspects of topic quality, such as topic distinctiveness and topic credibility nor consider the likelihood of the resulting models. 

With regards to the label-switching problem, which also affects LDA since it is inherently a mixture model, \cite{stephens2000dealing,celeux1998bayesian,stephens1997bayesian} developed relabelling algorithms to perform a k-means type clustering of the MCMC samples. \cite{hastie2015sampling} followed a k-medoid strategy to obtain an optimal partition that takes advantage of the whole MCMC output rather than taking a maximum a posteriori partition. Other relabelling strategies consider label invariant loss functions \citep{celeux2000computational,hurn2003estimating}, identifiability constraints \citep{mclachlan2019finite}, and probabilistic relabelling \citep{jasra2005markov,sperrin2010probabilistic}. Note that these techniques assume that topics are present (but with switched labels) across samples. Thus, we cannot use relabelling techniques to summarize topic models, since topics may (dis)appear across a Markov chain. Instead, we propose a methodology to group topics using similarity measures.

\section{Latent Dirichlet Allocation\label{sec:LDA}}

Here, we interpret Latent Dirichlet Allocation (LDA) \citep{blei2003latent} in terms of retail data, where transactions are interpreted as \textit{bags of products}. This is a natural assumption of in-store transactions where products are registered without an inherited order. In addition, transactions are assumed to be independent and exchangeable, so metadata such as timestamps and coordinate location are disregarded.
Within the LDA framework, transactions are represented as mixtures over a finite number of topics \(K\) and topics are distributions over products from a fixed product assortment of size \(V\). More formally, LDA is a generative process in which the topics \(\Phi=[\phi_1,...,\phi_K]\) are sampled from a Dirichlet distribution governed by hyperparameters \(\boldsymbol{\beta} = [\beta_1, ..., \beta_V]\) and the topical mixtures \(\Theta=[\theta_1,...,\theta_D]\) are sampled from a Dirichlet distribution governed by hyperparameters \(\boldsymbol{\alpha}= [\alpha_1, ..., \alpha_K]\). For each transaction (equivalent to a basket) $d$, product $i$ is sampled through a two-step process. First, a topic assignment \(z_{i,d}\) is chosen from the transaction-specific topical mixture \(\theta_d\). Second, a product is sampled from the assigned topic \(\phi_{z_{i,d}} \). Mathematically,

\begin{equation}
  \label{eq:LDAgenerativemodel}
  \begin{aligned}
    \phi_k &\sim Dirichlet(\boldsymbol{\beta})\\
    \theta_d &\sim Dirichlet(\boldsymbol{\alpha})\\
    z_{i,d} | \theta_d &\sim Multinomial(\theta_{d}) \\
    w_{i,d} | \phi_{z_{i,d} } &\sim Multinomial(\phi_{z_{i,d} }).
  \end{aligned}
\end{equation}

The data then correspond to the observed set of products $w_{i,d}$ within each transaction $d$. The posterior distribution of the topic distributions \(\Phi\) and topical mixtures \(\Theta\) are given by the posterior conditional probability:
 
\begin{equation}
  \label{eq:LDAPosteriorDistribution}
  P(\Phi, \Theta,\textbf{z}|\textbf{w},\boldsymbol{\alpha},\boldsymbol{\beta}) = \frac{P(\Phi, \Theta,\textbf{z},\textbf{w}|\boldsymbol{\alpha},\boldsymbol{\beta})}{P(\textbf{w}|\boldsymbol{\alpha},\boldsymbol{\beta})},
\end{equation} 
where \(\textbf{z}\) and \(\textbf{w}\) are vectors of topic assignments and observable products, respectively. In this paper, we use the collapsed Gibbs sampling algorithm \citep{griffiths2004finding} to sample from the posterior distribution and learn topic distributions since this method has shown advantages on computational implementation, memory, and speed.

In this paper, we used LDA with symmetric Dirichlet priors governed by a scalar concentration parameter and a uniform base measure, so that topics are equally likely a priori. \cite{wallach2009rethinking} showed that an optimized asymmetric Dirichlet prior over topical mixtures improves model generalization and topic interpretability by capturing highly frequent terms in a few topics. However, we empirically found that LDA with an asymmetric prior may lead to poor convergence of the Gibbs sampler in the context of our application. Finally, here we also assume that the number of topics is fixed and known a priori, but the proposed method can also be applied to the Hierarchical Dirichlet Process \citep{teh2005sharing}.

\subsection{Gibbs sampling}

The Gibbs sampling algorithm starts with a random initialization of topic assignments \(\mathbf{z}\) to values \(1,2,..., K\). In each iteration, topic assignments are sampled from the full conditional distribution, defined as:

\begin{equation}
  \label{eq:LDAFullCondDistribution}
  p(z_i=k|\mathbf{z}_{-i},\mathbf{w})\propto
  \dfrac{N_{k,v}^{-i}+\beta_v}
  {N_{k}^{-i}+\beta}
  \dfrac{N_{d,k}^{-i}+\alpha_k}
  {N_{d}^{-i}+\alpha},
\end{equation}
where the notation \(N^{-i}\) is a count that does not include the current assignment of \(z_i\). \(N_{k,v}\) is the number of assignments of product \(v\) to topic \(k\). \(N_{d,k}\) is the number of assignments of topic \(k\) in transaction \(d\). \(N_{k}\) is the total number of assignments of topic \(k\). \(N_{d}\) is the size of transaction \(d\). \(\alpha = \sum_k^K\alpha_k\) and \(\beta = \sum_v^V\beta_v\). This full conditional distribution can be interpreted as the product of the probability of the product \(v\) under topic \(k\) and the probability of topic \(k\) under the current topic distribution for transaction \(d\). Consequently, the probability of assigning a topic to any particular product in a transaction will be increased once many products of the same type have been assigned to the topic and the topic has been assigned several times to the transaction.

After a burn-in period, states of the Markov chain (topic assignments) are recorded with an appropriate lag to ensure low autocorrelation between samples. For a single sample $s$, \(\Phi\) and \(\Theta\) are estimated from the counts of topic assignments and Dirichlet parameters by their conditional posterior means:
\begin{equation} {\hat{\phi}}^{s}_{k,v} = E(\phi^s_{k,v}|\textbf{z}^s,\boldsymbol{\beta})= \dfrac{N_{k,v}^s+\beta_v^s}{N_{k}^s + \beta^s},\;\;k = 1\ldots K, v=1\ldots V,
\end{equation}
\begin{equation} {\hat{\theta}}^{s}_{d,k}= E(\theta^s_{d,k}|\textbf{z}^s,\boldsymbol{\alpha}) = \dfrac{N_{d,k}^s+\alpha_k^s}{N_{d}^s + \alpha^s},\;\;d=1\ldots D, k = 1\ldots K.
\end{equation}
%which are the predictive distributions over new products and new topics conditioned on \(\mathbf{w}\) and \(\mathbf{z}\).

\section{Topic model evaluation\label{sec:evaluation}}

Topic model evaluation is typically based on model fit metrics such as held-out-likelihood or perplexity \citep{wallach2009evaluation,buntine2009estimating}, which assess the generalization capability of the model by computing the model likelihood on unseen data. However, the LDA likelihood may lead to topic models with less semantically meaningful topics according to human annotators \citep{chang2009reading}. The evaluation of topic models should therefore not be exclusively based on likelihood metrics, but also include topic quality metrics such as topic coherence, topic distinctiveness, and topic credibility.

In this section, we summarise metrics of model generalization, topic coherence, and introduce metrics for topic distinctiveness and topic credibility. These four metrics will be used to evaluate topic models throughout this paper.

\subsection{Model generalization}

Model fit metrics such as perplexity or held-out-likelihood of unseen documents (transactions) estimate the model's capability for generalization or predictive power. Perplexity is a measurement of how well the probability model predicts a sample of unseen (or seen) data. A lower perplexity indicates the topic model is better at predicting the sample. Mathematically,

\begin{equation}
  \label{eq:Perplexity}
  \textrm{Perplexity}=-\frac{\log P(\mathbf{w}^\prime|\Phi,\boldsymbol{\alpha})}{N^\prime},
\end{equation}
where \(\mathbf{w}^\prime\) is a set of unseen products in a document, \(N^\prime\) is the number of products in \(\mathbf{w}^\prime\), \(\Phi=[\phi_1, \phi_2,\ldots,\phi_K]\) is a posterior estimate or draw of topics and \(\boldsymbol{\alpha}\) is the posterior estimate or draw of the Dirichlet hyperparameters.

Computing the log-likelihood of a topic model on unseen data is an intractable task. Several estimation methods are described in \citep{wallach2009evaluation,buntine2009estimating}. In this paper, we use the left-to-right algorithm with 30 particles to approximate the log-likelihood on held-out documents \citep{wallach2009evaluation, wallach2008structured}. The left-to-right algorithm breaks the problem of approximating the log-likelihood of one document (transaction) in a series of parts, where each part is associated to the probability of observing one term (product) given the previously observed terms. The likelihood of each term is approximated using an approach inspired by sequential Monte Carlo methods, where topic assignments are resampled for the previously observed terms to simulate topical mixtures over observed terms. The likelihood is given by the summation over topics of the product between the probability of the topic in the document and the probability of the term under the topic distribution. This procedure is repeated for a number of iterations (particles) and the likelihood of the term is given by averaging the per-particle likelihood.

\subsection{Topic coherence}

A topic is said to be coherent when its most likely terms can be interpreted and associated with a single semantic concept \citep{newman2010automatic}. For instance, `a bag of egg noodles', `a package of prepared stir fry', and `a sachet of Chinese stir fry' sauce are items that can be easily associated with the topic of `Asian stir fry'. On the other hand, a non-coherent topic highlights products that do not seem to fulfil a particular customer need. For example, `a bag of egg noodles', `a bunch of bananas', and `a lemon cake' are items that together do not convey a clear purpose.

Human judgement on topic coherence tends to correlate with metrics of product co-occurrence such as the Pointwise Mutual Information (PMI) and Normalized Pointwise Mutual Information (NPMI) \citep{bouma2009normalized} shown in \cite{newman2010automatic,lau2014machine}. PMI measures the probability of seeing two products within the same topic in comparison to the probability of seeing them individually. NPMI standardizes PMI, providing a score in the range of \([-1,1]\). NPMI towards \(1\) corresponds to high co-occurrence. 

\begin{equation}
  \label{eq:PMI}
  \textrm{PMI}(w_i, w_j)=\log\Big(\frac{P(w_i,w_j)}{P(w_i)P(w_j)}\Big); \quad i
  \neq j,\;\; 1\leq i,j\leq 15.
\end{equation}

\begin{equation}
  \label{eq:NPMI}
  \textrm{NPMI}(w_i, w_j)=\frac{\textrm{PMI}(w_i, w_j)}{-\log P(w_i,w_j)}; \quad i \neq j,\;\; 1\leq i,j\leq 15.
\end{equation}

Scores are calculated over pair combinations of the 15 most probable products, following \cite{blei2003latent,griffiths2004finding,steyvers2007probabilistic,chang2009reading,newman2010automatic,chaney2012visualizing}, 
Alternatively, products can be selected using distributional transformations \cite{taddy2012estimation,chuang2012termite,sievert2014ldavis} which highlight less frequent but topic-wise unique products. However, transformations may select terms with low probabilities under the topic distribution. 

The coherence measure of a single topic is given by the average of the NPMI scores. For simplicity, we will refer to this measure as NPMI. Here, we focus on NPMI since it has been shown to have a higher correlation with the human evaluation of topic coherence than PMI \citep{lau2014machine}. 

\subsection{Topic distinctiveness}
  
Topic distinctiveness refers to the semantic dissimilarity of one topic in comparison to the topics of the same sample. For instance, `a bottle of sparkling water hint apple', `a bottle of sparkling water hint grape', and `a bottle of sparkling water hint orange' are items that are interpreted as the topic of `flavoured sparkling water'. This topic and the `Asian stir fry' topic are distinctive from each other. If a topic in the posterior sample is characterized by `a bottle of sparkling water hint lemon', `a bottle of sparkling water hint mango' and `a bottle of sparkling water hint lime', it is interpreted as non-distinctive from the `flavoured sparkling water' since both topics exhibit the same theme. 

Several measures have been used to identify similar topics: KL-divergence  \citep{li2006pachinko,wang2009mining,newman2009distributed}, the average log odds ratio \citep{chaney2012visualizing}, the cosine distance \citep{ramage2009labeled,he2009detecting,chuang2015topiccheck, xing2018diagnosing}. \cite{aletras2014measuring,xing2018diagnosing} showed that cosine distance outperforms other distributional similarity measures, such as KL-divergence, Jensen Shannon Divergence, Euclidean distance, Jaccard similarity, according to human judgment on topic similarity. Thus, we define the distinctiveness of a topic $\phi^t_i$ of posterior draw $t$ as the minimum of the cosine distances between the topic and the other topics  $\Phi^t\setminus \phi^t_{i}$ within the same posterior sample, denoted by :

\begin{equation}
  \label{eq:topicDist}
  \textrm{CD}_{\min}\left(\phi^t_i, \Phi^t\setminus \phi^t_i\right) = \min \big[ \textrm{CD}(\phi^t_i,\phi^t_1),...,\textrm{CD}(\phi^t_i,\phi^t_{i-1}),\textrm{CD}(\phi^t_i,\phi^t_{i+1}),..., \textrm{CD}(\phi^t_i,\phi^t_K)\big], 
\end{equation}
where
\begin{equation}
  \label{eq:cosineDist}
\textrm{CD}\left(\phi_i,\phi_j\right)= 1-\frac{\phi_i \cdot\ \phi_j}{\parallel \phi_i \parallel \parallel \phi_j \parallel}.
\end{equation}
Cosine distance between topics measures a slightly different aspect of a topic compared to the model likelihood, and thus the model may warrant the existence of two similar topics in terms of cosine distance, showing a low minimum distance. The distinctiveness of a set of topics in a posterior sample is given by the average per-topic distinctiveness.

\subsection{Topic credibility}
When comparing different LDA posterior draws, topics may appear and disappear as a result of posterior uncertainty, which negatively affects practitioners' confidence in the method. While topic distinctiveness within the same posterior sample is good, the high cosine distance of topic $\phi^t_i$ with all topics $\Phi^s$ in posterior draw $s\neq t$ indicates uncertainty about $\phi^t_i$. To measure topic credibility of topic $\phi_i^t$ in posterior draw $t$, we compute the average maximum cosine similarity between $\phi_i^t$ and all topics within posterior draw $\Phi^s$, for $s\neq t$, and average across all posterior draws $s\neq t$. If a topic is highly credible, then we expect a very similar topic to appear in every single posterior draw, hence the average cosine similarity will be high. Note here that we are using cosine similarity, rather than cosine distance, to capture topic credibility. 

In other words,

\begin{equation}
  \label{eq:MaxcosineSim}
  \textrm{CS}_{\max}\left(\phi^t_i, \Phi^s\right) = \max \big[ \textrm{CS}(\phi^t_i,\phi^s_1),..., \textrm{CS}(\phi^t_i,\phi^s_K)\big], 
\end{equation}
where
\begin{equation}
  \label{eq:cosineSimi}
\textrm{CS}\left(\phi_i,\phi_j\right)= \frac{\phi_i \cdot\ \phi_j}{\parallel \phi_i \parallel \parallel \phi_j \parallel}.
\end{equation}
Averaging across all other posterior draws,
\begin{equation}
  \label{eq:topicCred}
 \overline{\textrm{CS}_{\max}}(\phi^t_{i}, \Phi^{1:S})= \frac{\sum_{s\neq t}  \textrm{CS}_{\max}\left(\phi^t_i,\Phi_{\cdot}^{s}\right)}{S-1}.
\end{equation}
A large average of the maximum similarities (i.e., minimum distances) across samples indicates that the topic appears with high similarity across posterior samples. The credibility of a set of topics is given by the average per-topic credibility.

\section{Posterior summary of topic distributions \label{sec:LDAclustering}}

Here we introduce a methodology that aims to summarise the posterior distribution of a topic model by quantifying the recurrence of topic modes across posterior samples. Recurrent topics tend to appear several times across LDA posterior draws, showing higher credibility. To group topics across samples that represent the same theme, we use a hierarchical clustering approach that retrieves clusters of topical similarity. The resulting clusters are used to quantify topic posterior recurrence of a clustered topic, which is ultimately used to identify and filter out topics of high uncertainty. 

\subsection{Hierarchical clustering}

Agglomerative hierarchical clustering (AHC) is a widely used statistical method that groups units according to their similarity, following a bottom-up merging strategy. The algorithm starts with as many clusters as input topics, and at each step, the AHC merges the pair of clusters with the smallest distance. AHC finishes when all the units are aggregated in a single cluster or when the distance among clusters is larger than a fixed threshold. AHC does not require the user to fix the number of clusters a priori; instead, the clustering dendrogram can be `cut' at a user's desired level, potentially informed by domain knowledge.

We use the AHC algorithm to aggregate and fuse topics from multiple posterior samples.  To quantify cluster similarity, we use CD and the average linkage method. We opt for CD since it has outperformed correlation on human evaluation of topic similarity \citep{aletras2014measuring} and human rating of posterior variability \citep{xing2018diagnosing}. We opt for the \textit{average} linkage method since, empirically, it has worked better than \textit{single} and \textit{complete} linkage methods, i.e., single linkage tended to create an extremely large cluster of low coherence, and complete linkage tended to create clusters of low distinctiveness. However, we slightly modify the algorithm to merge only topics that come from different posterior samples and whose cosine distance is lower than a user-specified threshold. In this manner, we avoid merging topics that belong to the same posterior sample or that differ to such a large extent that merging them is meaningless.

\subsection{Recurrent topics}

The AHC retrieves a collection of clusters \(C_1, ..., C_N\), which are represented by a \textit{clustered topic} \(\overline{\phi_k}\) with a \textit{cluster size} \(\vert C_k \vert\), where \(k=1,...N\). The clustered topic is the average distribution of the topics that share the same membership. The cluster size is the number of members, e.g., clustering 100 identical posterior samples of 50 topics would retrieve 50 clusters of 100 members each. The cluster size also represents the uncertainty related to the clustered topic. For instance, a cluster of size one indicates that its associated topic does not reappear in other posterior samples. On the other hand, a recurrent topic would be associated with a cluster with large cluster size, indicating that the topic consistently reappears across multiple samples. Thus, we measure the recurrence of a topic by its cluster size:

\begin{equation}
  \label{eq:ClusteredDist}
  \begin{aligned}
    \textrm{recurrence}(\overline{\phi}_{i}) = {\lvert C_i \rvert}.
  \end{aligned}
\end{equation}

Then, subsets of clustered topics filtered by their recurrence are evaluated to identify a subset of clustered topics with high credibility. As we will show in the next section, cluster size as a measure of topic recurrence leads to subsets of better topic quality.

\section{Application to grocery retail data\label{sec:groceryuserstudy}}

We apply topic models in the domain of the grocery retail industry, where topics are distributions over a fixed assortment of products and transactions are described as mixtures of topics. We analyse grocery transactions from a major retailer in the UK. Transactions are sampled randomly randomly, covering 100 nationwide superstores between September 2017 and August 2018. The training data set contains 36 thousand transactions and a total of 392,840 products and the test data set contains 36 hundred transactions and a total of 38,621 products. Transactions contain at least 3 products and 10 products on average. The product assortment contains 10,000 products which are the most monthly frequent, ensuring the selection of seasonal and non-seasonal products. We count unique products in transactions, disregarding the quantities of repetitive products. For instance, 5 loose bananas count as 1 product (loose banana). We do not use an equivalent of stop words list (highly frequent terms), as we consider that every product or combination of them tell different customer needs. We disregard transactions with fewer than 3 products assuming that smaller transactions do not have enough products to exhibit a customer need. No personal customer data were used for this research.

\subsection{Human judgement on interpretability and similarity of topics}

To aid interpretation of topics within the context of the application, meaningful NPMI and cosine similarity thresholds need to be set. To this end, we carried out a user study to collect human judgement on the interpretability of individual topics and the similarity between pairs of topics and, ultimately, set empirical thresholds driven by users' interpretations. Experts from a leading data science company specializing in retail analytics participated in the user study.

Users were asked to evaluate topics using a discrete scale from 1 to 5. For similarity between a pair of topics, a score of 1 refers to highly different topics, and a score of 5 refers to highly similar topics. For interpretability, a score of 1 refers to highly incoherent topics, and a score of 5 refers to highly coherent topics. Topics were obtained from \(25, 50, 75, 100, 125,150\)-topic LDA with hyper-parameters \(\alpha=[0.1, 0.01]\) and \(\beta=[0.01,0.001]\). The range in the number of topics corresponds to an initial belief of having no less than 25 topics and no more than 150 topics. Topics were represented by the top 10 most probable products. 189 and 935 evaluations for topic distinctiveness and topic coherence were collected, respectively.

\begin{figure}[H]
  \centering
  \begin{subfigure}[b]{0.48\textwidth}
    \centering
    \caption{Interpretability}
    \includegraphics[width=0.95\linewidth]{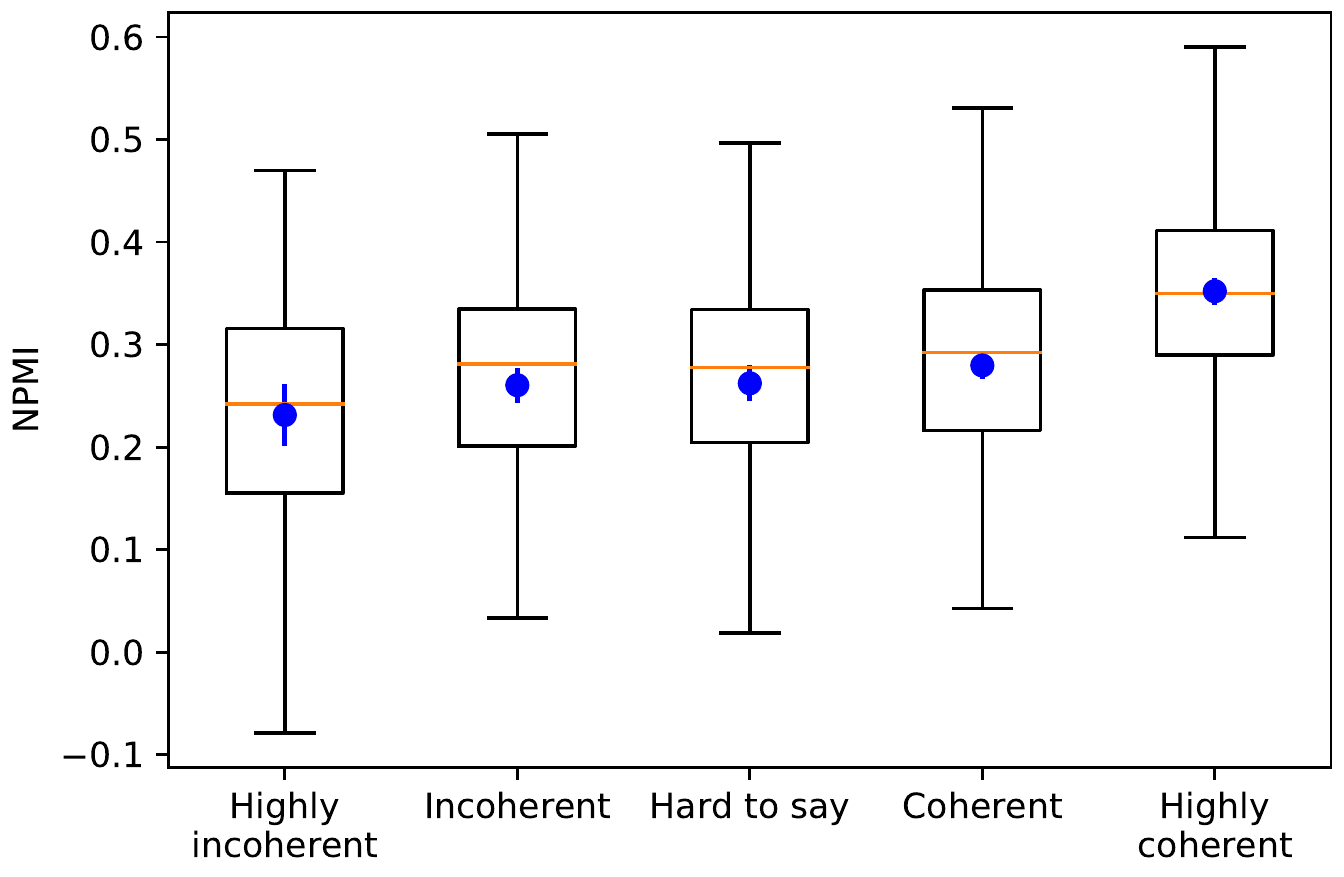}
    \label{npmi_survey}
  \end{subfigure} %
  \begin{subfigure}[b]{0.48\textwidth}
    \centering
    \caption{Similarity}
    \includegraphics[width=0.95\linewidth]{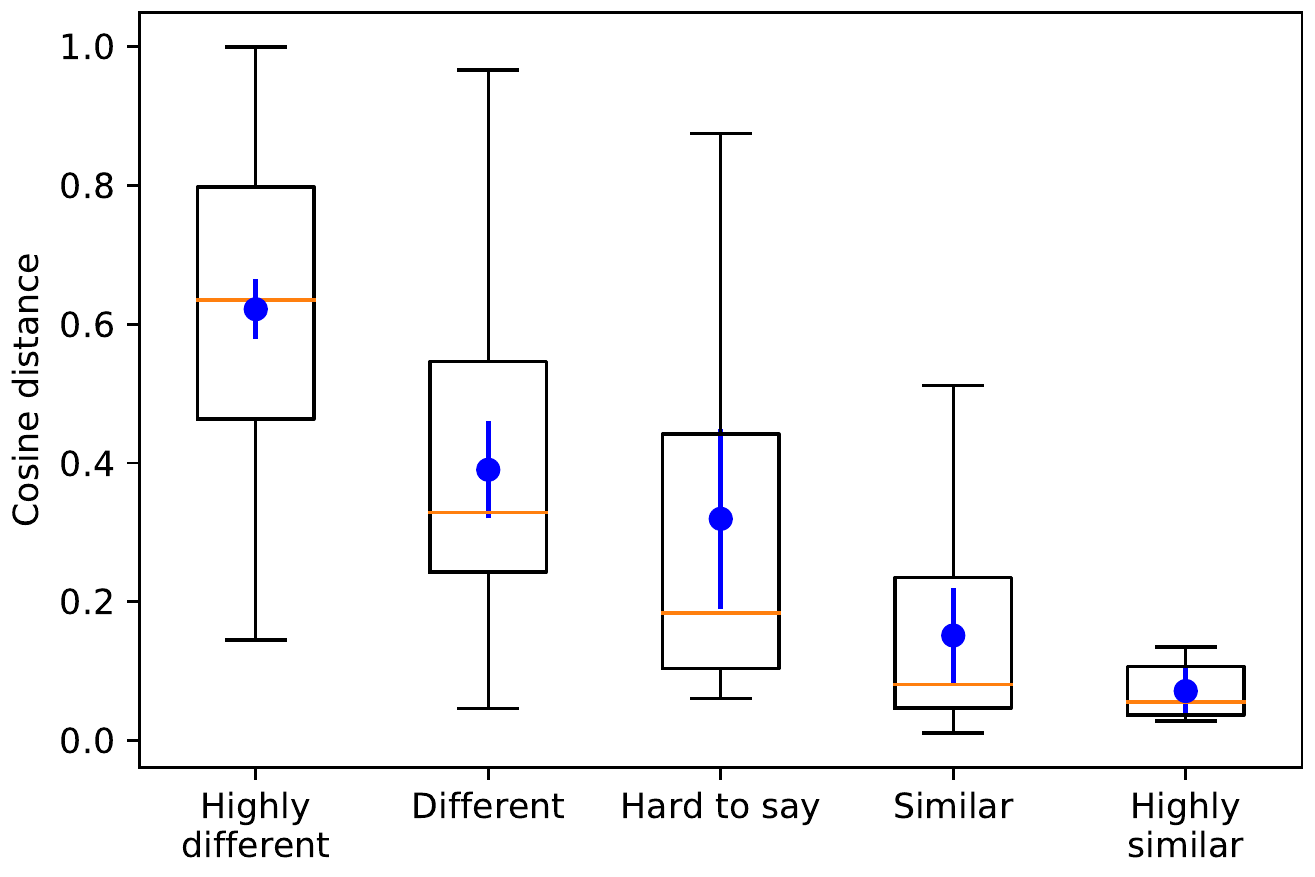}
    \label{sim_survey}
  \end{subfigure}%
  \caption{Human evaluation on interpretability of individual topics and similarity between pairs of topics. Figure \ref{npmi_survey} shows coherence scores against topic NPMI. Figure \ref{sim_survey} shows similarity scores against the cosine distance between compared topic distributions. Blue error bars show means and confidence intervals for the means. Interpreting results, a \(\textrm{CD} \leq 0.1\) indicates high similarity while \(\textrm{CD} \geq 0.5\) indicates high dissimilarity. It is also observed \(\textrm{NPMI} \leq 0\) responds to incoherent topics and \(\textrm{NPMI} \geq 0.5\) responds to highly coherent topics.}%
\end{figure}

Figure \ref{npmi_survey} compares human judgment on topic coherence against NPMI. Despite the subtle positive correlation, there is no clear boundary of NPMI that can precisely identify coherent topics. However, we observe that 100\% of topics with \(\textrm{NPMI} \leq 0\) were interpreted as highly incoherent, 65\% of topics with \(\textrm{NPMI} \geq 0.3\) were interpreted as coherent, and 96\% of topics with \(\textrm{NPMI} \geq 0.5\)  were interpreted as highly coherent. We use these interpretations to guide the interpretation of topic coherence in the next sections.

Figure \ref{sim_survey} compares human judgment on topic similarity against cosine distance. Unsurprisingly, the lower the cosine distance, the more similar the topic distributions are. We observe that 70\% of the pairs with \(\textrm{CD} \leq 0.1\) were interpreted as `Similar' or `Highly similar', and 95\% of pairs with \(\textrm{CD} \geq 0.5\) were interpreted as `Different' or `Highly different. While 38\% of pairs were interpreted as `Similar' or `Highly similar' when \(0.1 \geq \textrm{CD} \leq 0.3\), indicating some degree of topic similarity. Based on these results, we interpret topics with \(\textrm{CD} \leq 0.1\) as highly similar and with \(\textrm{CD} \geq 0.5\) as highly dissimilar. We use these thresholds to guide interpretations of topic distinctiveness and topic credibility.

\subsection{LDA performance} \label{LDAperformance}

We trained 5 LDA models with \(K=25, 50,100, 200, 400\) topics, with a symmetric Dirichlet hyperparameters \(\alpha_k=3/K\) and \(\beta_v=0.01\). Note that \(\sum_k alpha_k = 3\), which reflects the minimum transaction size. \(\beta_v=0.01\) is commonly used in the literature \citep{ mimno2011optimizing,newman2011improving}. For each model, 4 Markov chains are run for 50,000 iterations with a burn-in of 30,000 iterations; samples were recorded every 10 000 iterations obtaining 20 samples in total. As shown in Appendix \ref{mcmc_convergence}, convergence of the Markov chains is satisfactory.

LDA models are assessed on the four aforementioned quality aspects. Perplexity measures the generalization of a group of topics, thus it is calculated for an entire collected sample. The other evaluation metrics are calculated at the topic level (rather than at the sample level) to illustrate individual topic performance.

\begin{figure}[ht!]
    \centering
    \includegraphics[width=0.6\linewidth]{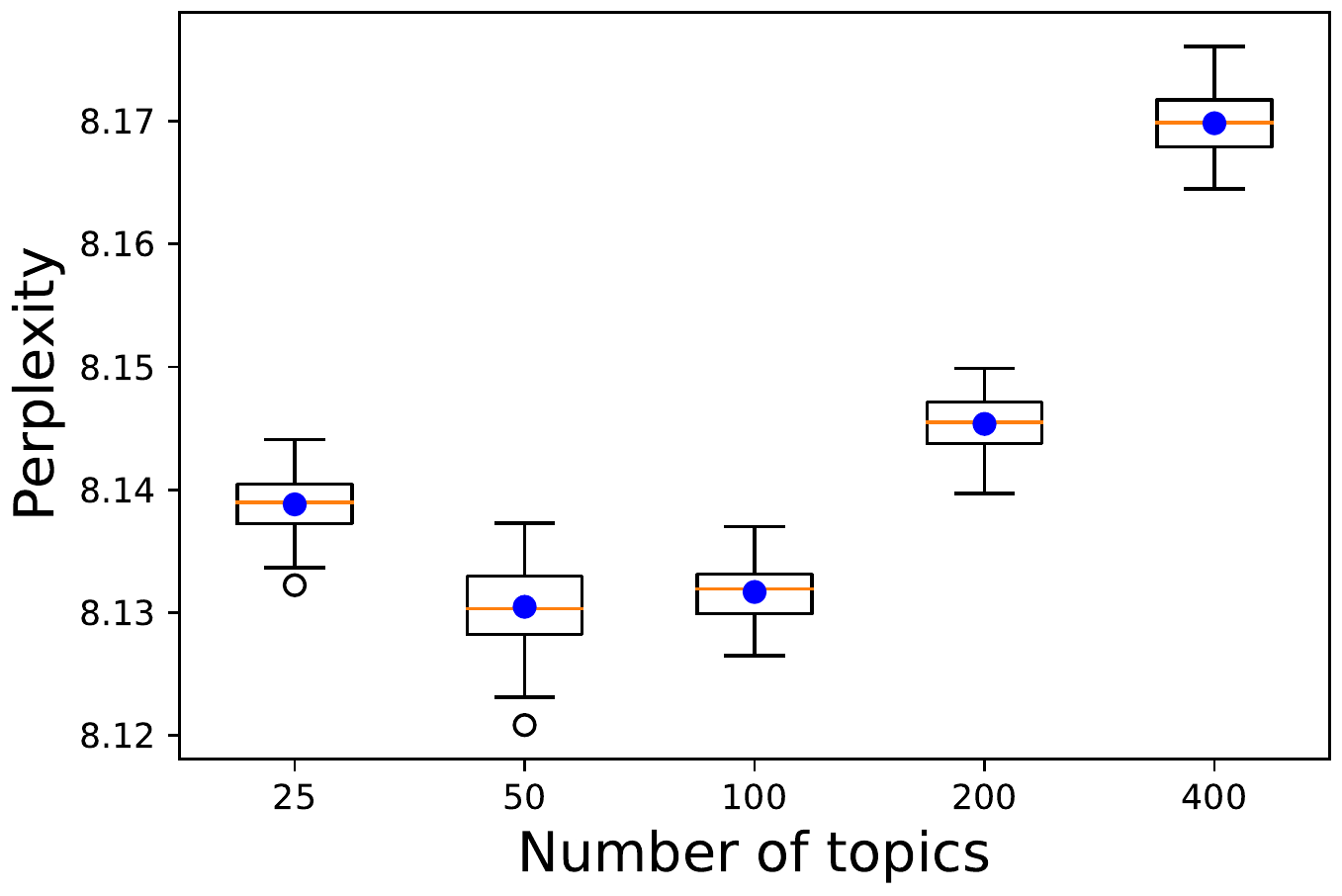}
\caption{The perplexity of LDA models with 25/50/100/200/400 topics. Each boxplot represents the perplexity distribution over the 20 samples. Green triangles indicate the average perplexity; standard errors are smaller than the marker size.}\label{N_perp}
\end{figure}

\begin{figure}[H]
 \centering
\begin{subfigure}[b]{0.65\textwidth}
    \centering
    \caption{Coherence across LDA models}
    \includegraphics[width=0.9\linewidth]{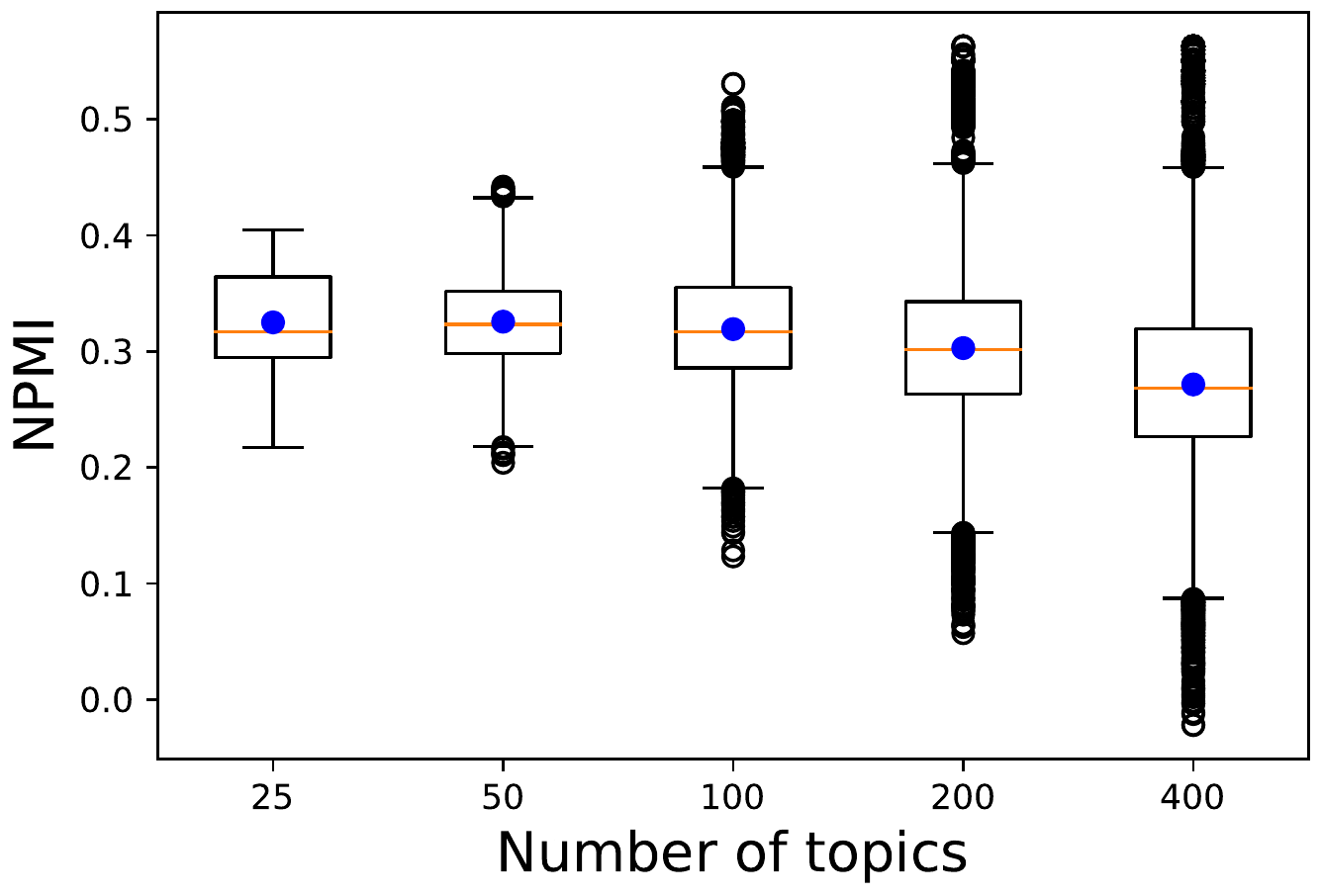}
    \label{N_npmi}
  \end{subfigure}%
  \begin{subfigure}[b]{0.28\textwidth}
    \centering
    \caption{Topics examples of low/high coherence}
    \includegraphics[width=1\textwidth]{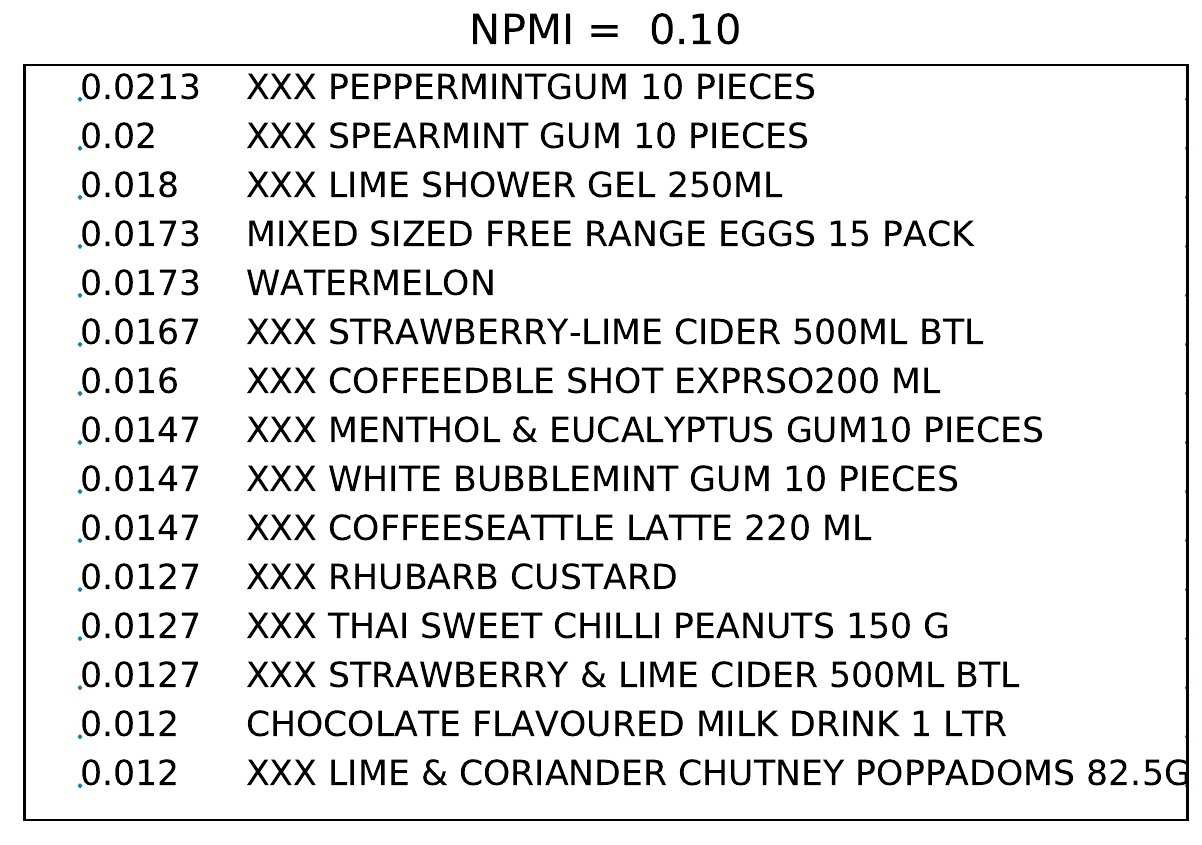} \\
\includegraphics[width=1\textwidth]{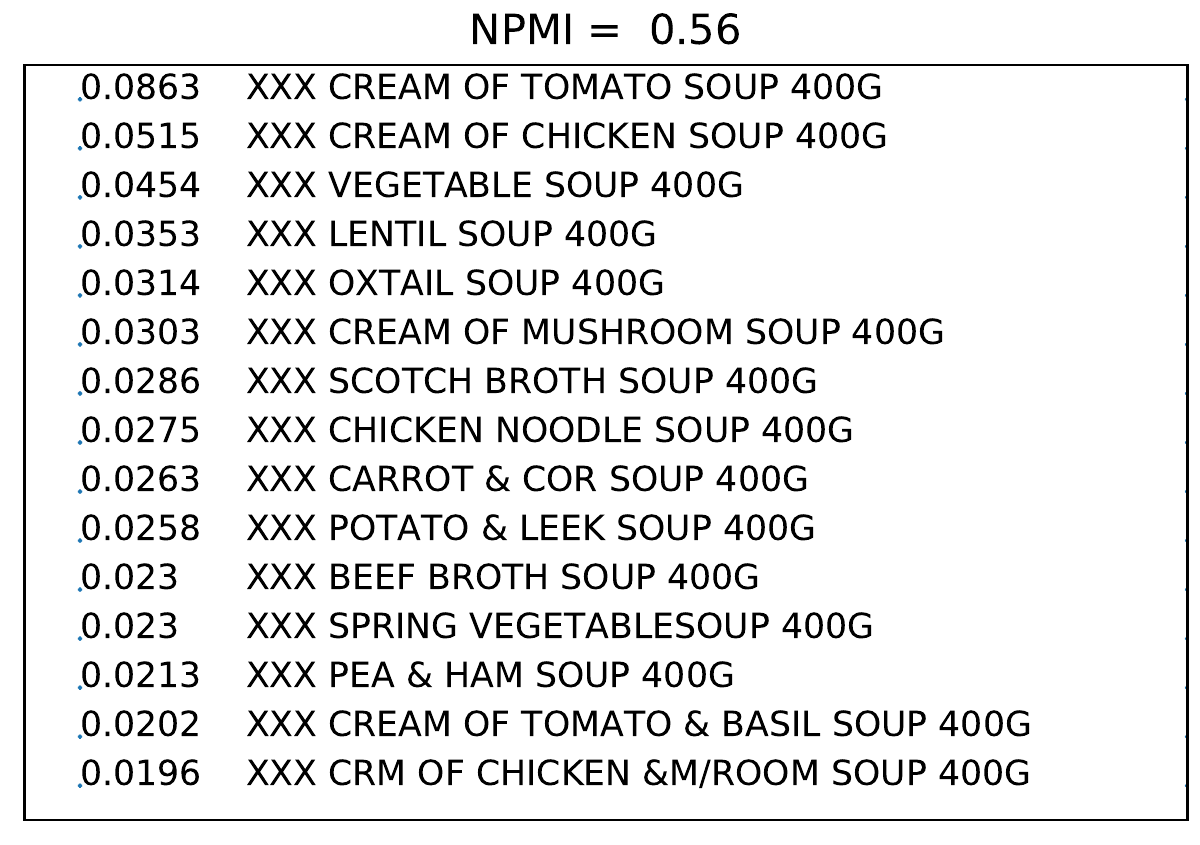}
    \label{coherence_examples}
  \end{subfigure}%
  \label{Coherence}
  \caption{(\ref{N_npmi}) Topic-specific NPMI of 25/50/100/200/400-topic LDA model. Green triangles indicate the average NPMI; standard errors are smaller than the marker size. (\ref{coherence_examples}) shows (top) a topic with low coherence, (bottom) a topic with high coherence. Topics are illustrated with the probability and description of the top 15 products. Brands have been replaced by XXX.}
\end{figure}

Figure \ref{N_perp} shows the perplexity performance of LDA models. LDA samples of 50 and 100 topics tend to have the best generalization capability. As observed in Figure \ref{N_npmi} posterior draws with 25 and 50 topics show larger average NPMI, however, there are no highly coherent topics ($\textrm{NMPI}>0.5$). The posterior draws with 100 to 400 topics show some highly coherent topics, but also show many less coherent topics with low NPMI values. In agreement with \cite{chang2009reading}, posterior samples with higher coherence do not necessarily have the best likelihood, which is the case of 25-topic LDA samples. Figure \ref{coherence_examples} illustrates two topics with low/high coherence. The top topic displays product descriptions that do not show a specific meaning, purpose, or customer need. On the other hand, the bottom topic shows the soup topic, composed of branded soup items that are frequently bought together due to promotional discounts.

\begin{figure}
 \centering
\begin{subfigure}[b]{0.65\textwidth}
    \centering
    \caption{Distinctiveness across LDA models}
    \includegraphics[width=0.9\linewidth]{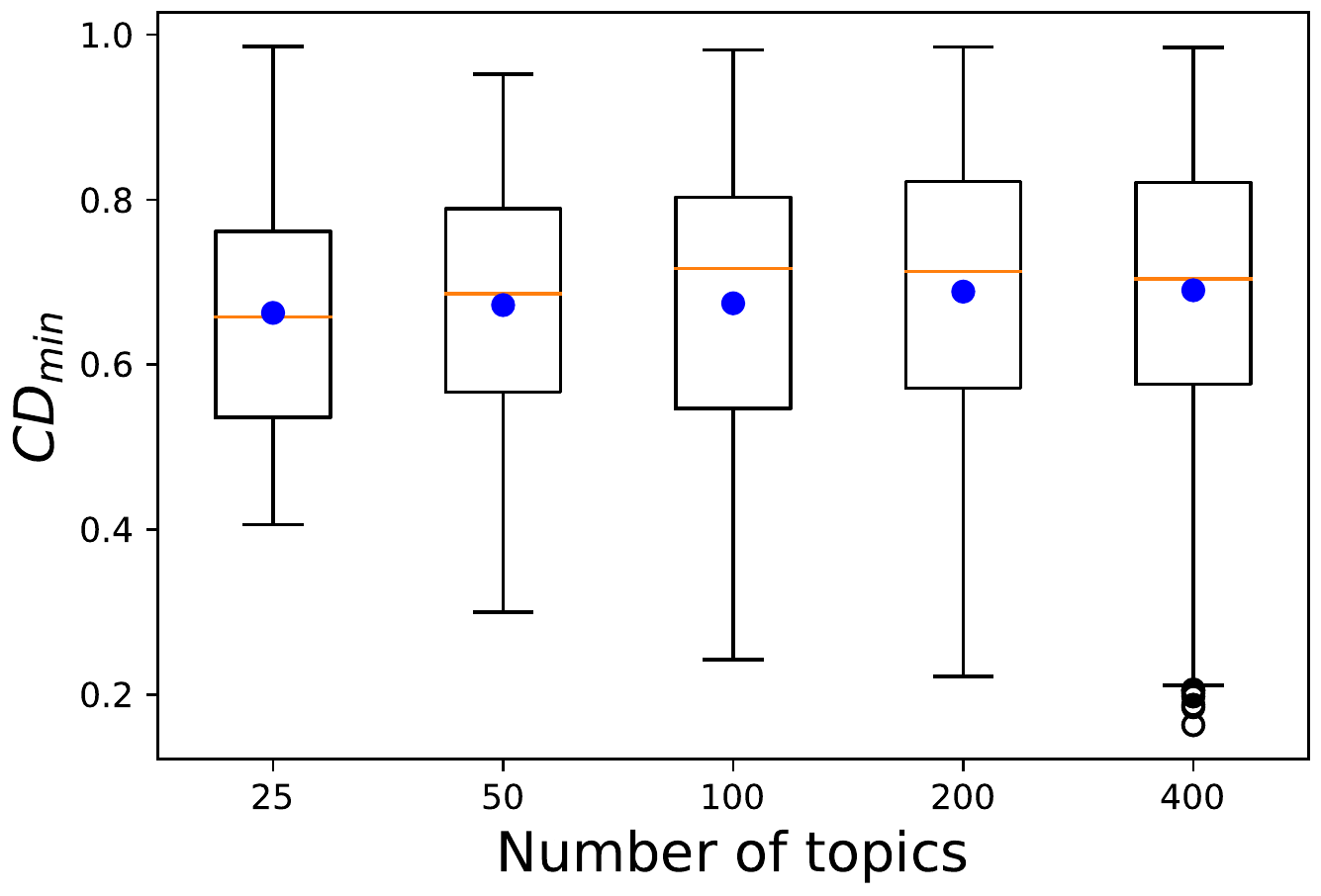}
    \label{N_Mincd}
  \end{subfigure}%
  \begin{subfigure}[b]{0.28\textwidth}
    \centering
    \caption{Topics examples of high similarity }
    \includegraphics[width=1\textwidth]{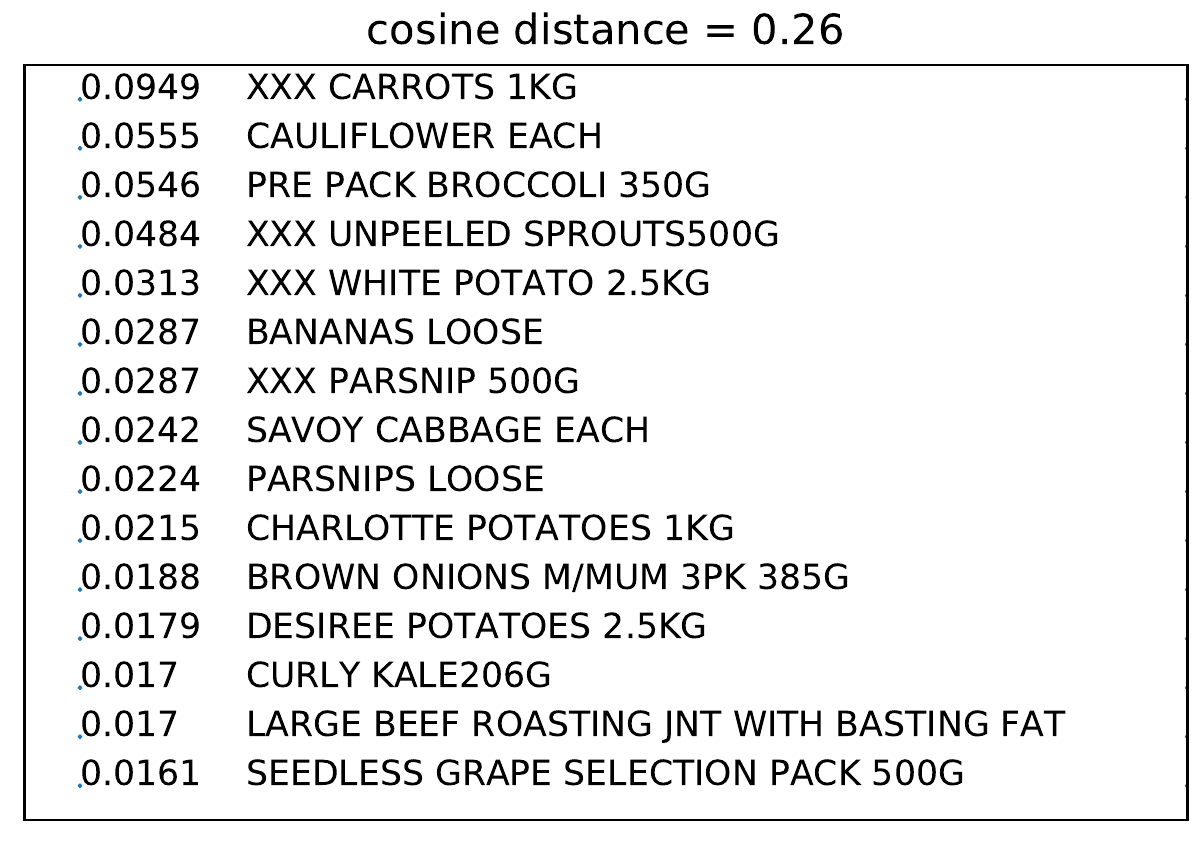}\\
\includegraphics[width=1\textwidth]{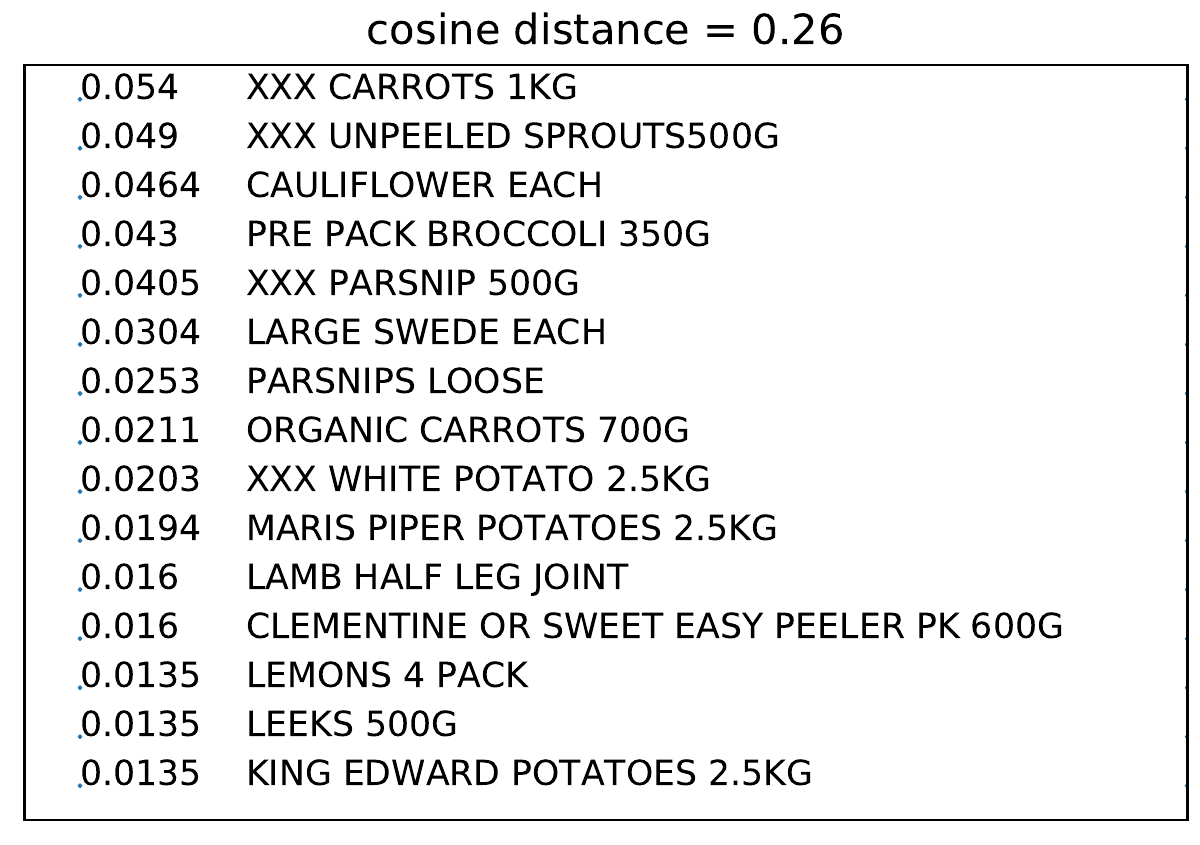}
    \label{distinctiveness_examples}
  \end{subfigure}%
  \caption{(\ref{N_Mincd}) Topic-specific minimum cosine distance (among topics of the same posterior draw). Green triangles indicate the average minimum cosine distance; standard errors are smaller than the marker size. (\ref{distinctiveness_examples}) shows two topics from a single Gibbs sample that show some similarity. Topics are illustrated with the probability and description of the top 15 products. Brands have been replaced by XXX.}
\end{figure}

In Figure \ref{N_Mincd}, we measure topic distinctiveness by computing the minimum cosine distance among topics of the same posterior draw. If two topics exhibit the same theme, and thereby similar distributions, then the cosine distance is close to 0. We observe that the majority of topics are highly distinct (\(\textrm{CD} \geq 0.5\)) within their posterior draw. However, as expected, the larger the model, the more topics with some degree of similarity (\(\textrm{CD} \leq 0.3\)) as seen in LDA models with 100 to 400 topics. Figure \ref{distinctiveness_examples} shows an example of two topics with some degree of similarity, both show collections of produce and red meat.  

\begin{figure}[H]
 \centering
\begin{subfigure}[b]{0.6\textwidth}
    \centering
    \caption{Credibility across LDA models}
    \includegraphics[width=0.99\linewidth]{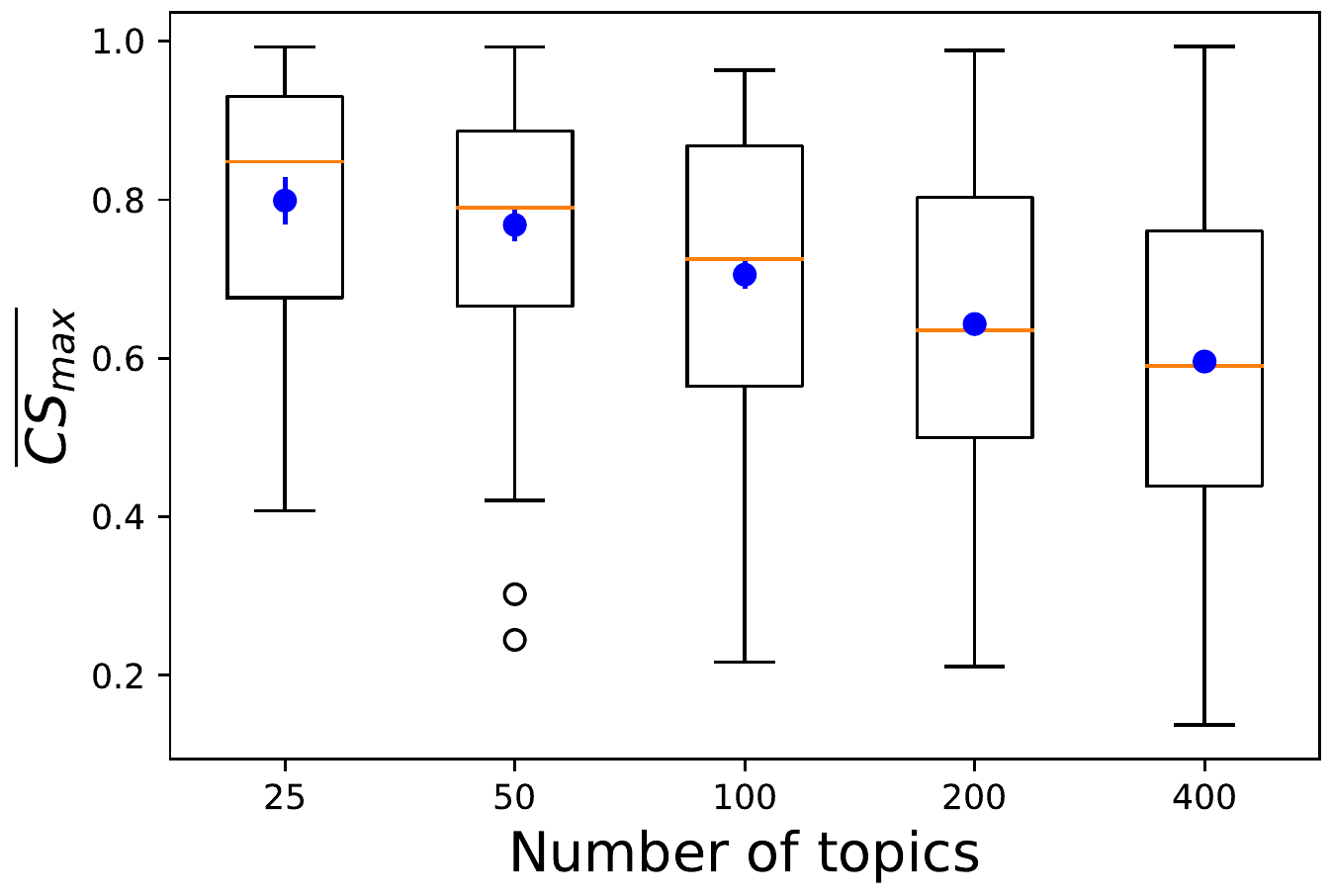}
    \label{N_CredAcChains}
  \end{subfigure}%
  \begin{subfigure}[b]{0.35\textwidth}
    \centering
    \caption{Topic similarity between two posterior draws.}
\includegraphics[width=1\textwidth]{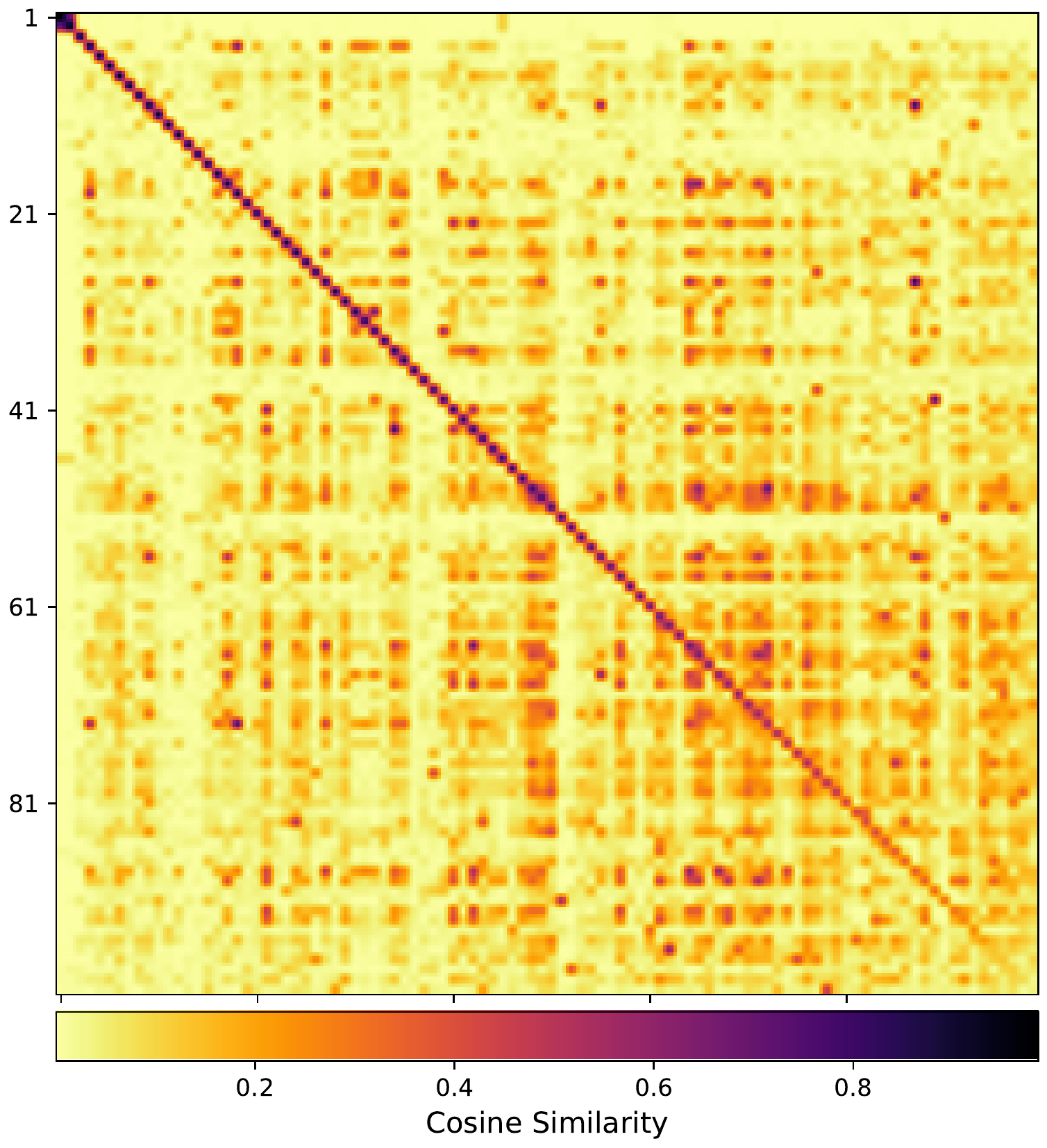}
    \label{credibility_example}
  \end{subfigure}%
    \caption{(\ref{N_CredAcChains}) Topic-specific average maximum cosine similarity. For each topic, the maximum cosine similarity is calculated over the topics of a different posterior draw. Then, the average is taken over all maximum values. When a topic is highly credible, it will frequently appear across posterior samples, thus the average maximum cosine similarity tends to 1. Conversely, if a topic is highly uncertain and it does not appear in other posterior samples, then the maximum cosine similarity for each sample would tend to zero. Green triangles indicate the mean; standard errors are smaller than the marker size. (\ref{credibility_example}) shows the cosine distance between topics of two posterior samples. Topics have been ordered using a greedy alignment algorithm that tries to find the best one-to-one topic correspondences.}
\end{figure}

In Figure \ref{N_CredAcChains}, we measure topic credibility by averaging the maximum cosine similarity between a topic and the topics from the remaining posterior samples, so for each topic and each sample, there is one maximum cosine similarity from each remaining posterior sample. If one topic constantly appears across samples, then the average maximum cosine similarity tends to 1. Vice-versa, if the topic is not part of other samples, then the maximum cosine similarity of each sample tends to 0, so does its average maximum cosine similarity. We observe 4\%/ 3\%/16\%/ 25\%/ 36\% of topics with \(\overline{\textrm{CS}_{\max}} \leq 0.5\), indicating that they did not reappear in other posterior samples with high similarity. Figure \ref{credibility_example} shows the cosine similarity matrix between two posterior LDA samples of 100 topics. Topics have been ordered using a greedy alignment algorithm that tries to find the best one-to-one topic correspondences as in \citep{rosen2010learning}. This plot indicates that around one-fifth of the topics do not appear with some similarity \(\textrm{CS} \geq 0.5\) in the other posterior draw. This implies that applying label-switching algorithms to resolve labelling for each posterior draw would inevitably `match-up' topics which are semantically dissimilar. Instead of averaging over distinct modes, our methodology (described in the next section) would report separate clusters, each with its own credibility, reflecting the frequency with which each mode appears.

\section{Clustering and selection of recurrent topics\label{sec:groceryclustering}}

In this section, we apply our methodology to summarize LDA posterior distributions and to quantify topic recurrence. We will show that topic recurrence can aid the selection of topics with better coherence, credibility and model generalization.

We conduct 3 experiments with LDA samples with 50, 100 and 200 topics. In each experiment, a bag of topics is formed from 20 samples that come from four separate Gibbs samplers. From each chain, samples are obtained after a burn-in period (30,000 iterations) and recorded every 5000 iterations to reduce autocorrelation. Computing perplexity is a computationally expensive. Thus, we do not record the evaluation metrics at each clustering step. Instead, we evaluate subsets of clustered topics obtained at different distance thresholds (cosine distance from 0 to 0.55 and every 0.05). We assume that topics with cosine distance \(\geq 0.55\) are too different, which would render cluster merging meaningless. Credibility is measured by comparing one clustering experiment against a second clustering experiment whose samples are recorded from four different Gibbs samplers. We do not further explore LDA samples with 25 and 400 topics, the former does not show a better variety of topic and the latter show worse perplexities. 

  \begin{figure}[H]
    \centering
    \includegraphics[width=1\textwidth]{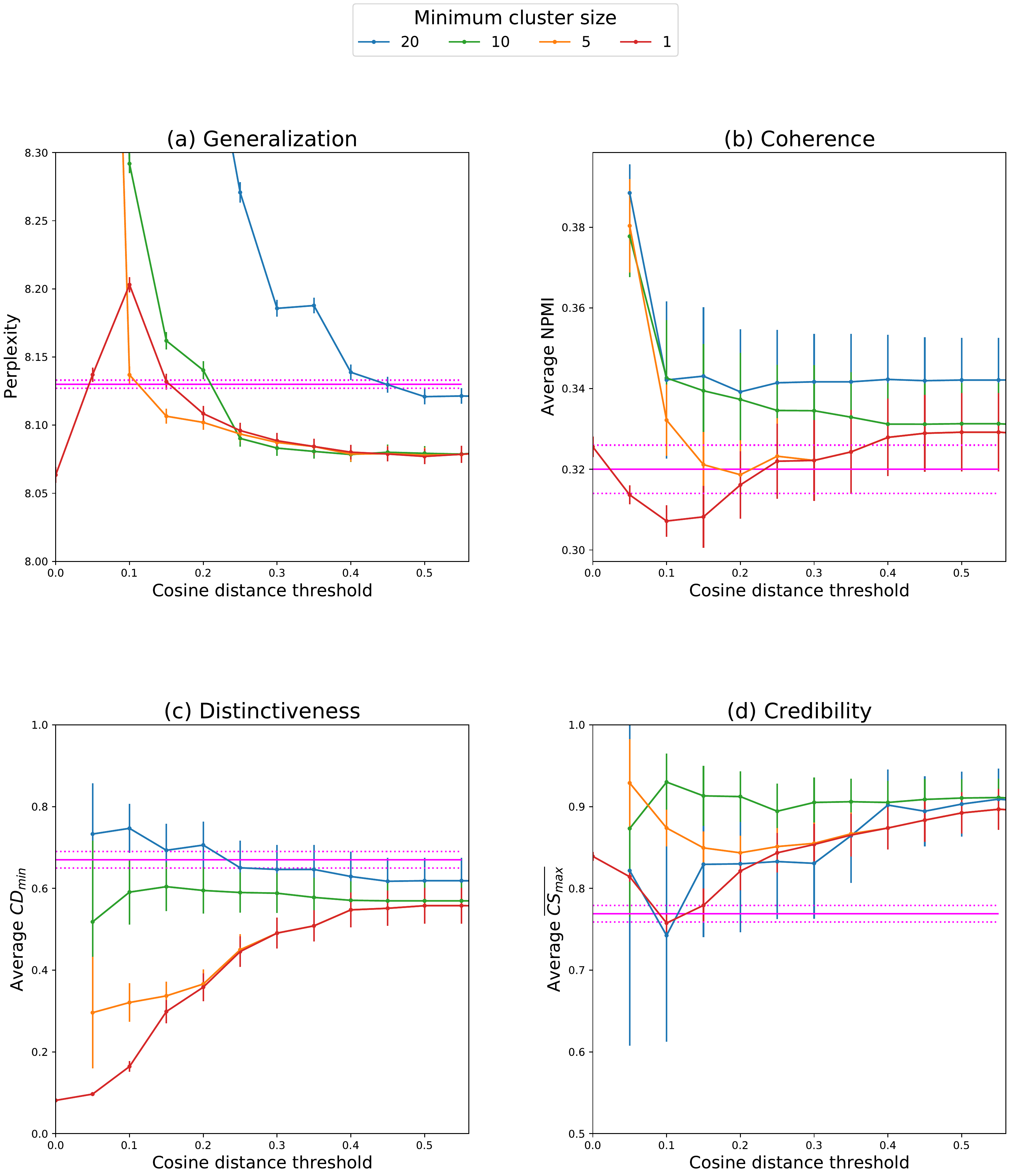}
    \caption{Subset evaluation using cosine distance (varying from 0 to 1 with increments of 0.05) and minimum cluster size (20, 10, 5 and 1). Clustered topics were obtained from clustering 20 samples of LDA with 50 topics. Vertical lines represent one standard error. Magenta lines show the average measures (\(\pm\) one standard error) of the LDA samples.}
\label{Evaluation}
  \end{figure}
    
\begin{figure}[H]
    \centering
    \includegraphics[width=0.58\textwidth]{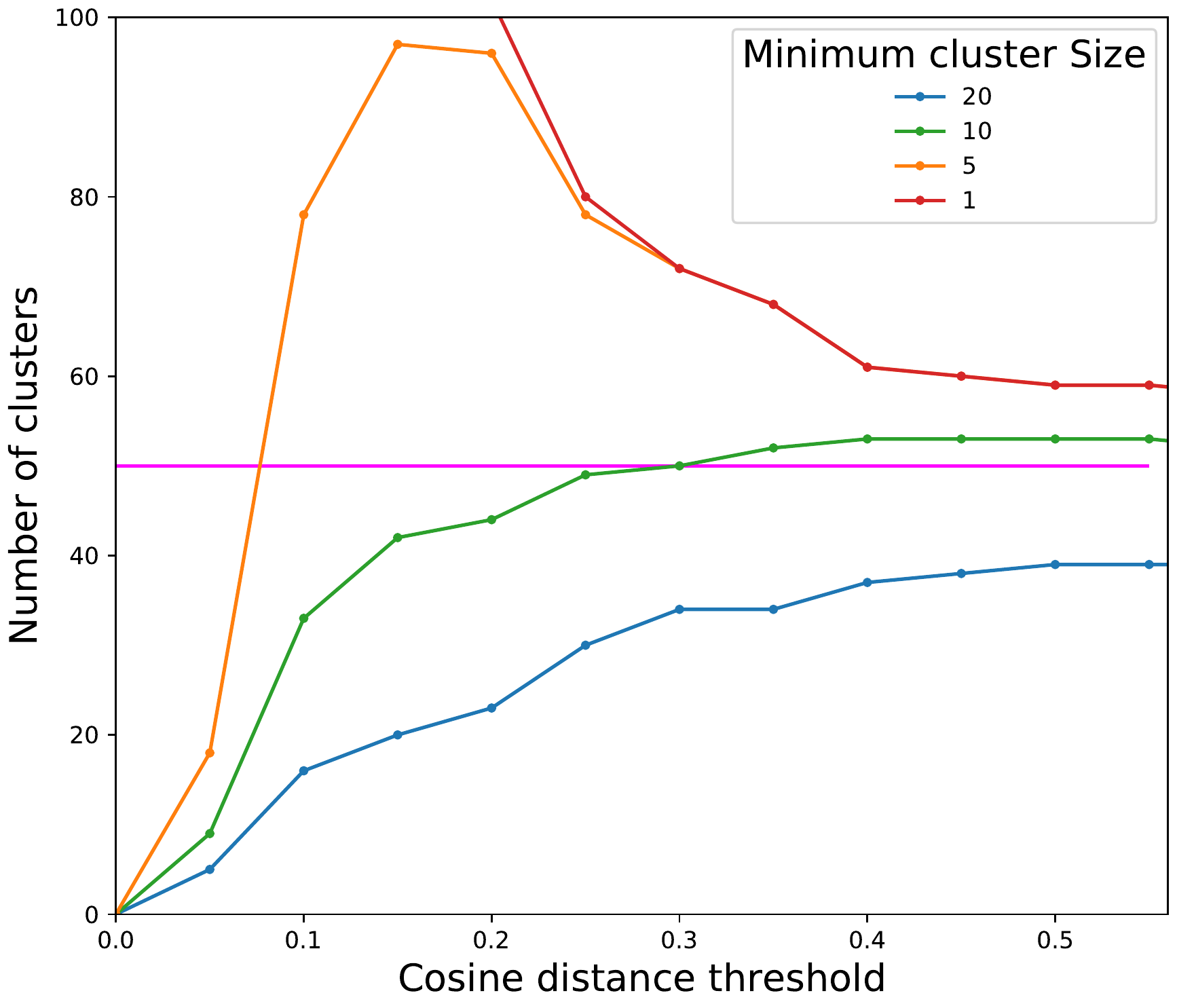}
    \caption{The number of clusters obtained at cosine distance (varying from 0 to 1 with increments of 0.05) and minimum cluster size (20, 10, 5 and 1). The magenta line shows the number of topics in the LDA samples. For visualization purposes, subsets larger than 100 clusters are not shown (subsets with minimum size 1 and \(CD<0.25\))}
     \label{ClusterNumber}
  \end{figure}

Figure \ref{Evaluation} shows the evaluation of subsets of clustered topics obtained from clustering 50-topic LDA samples at different levels of topic recurrence, when the minimum cluster size is 1, 5, 10, and 20, representing the 5\%, 25\%, 50\%, and 100\% of the samples. As observed in the perplexity plot (top left), the subset with the lowest perplexity is the one at minimum cluster size 1 and cosine distance 0, this is the original bag of 1000 topics before merging. This subset has the lowest performance in distinctiveness; thereby, using this subset is inefficient as it contains too many repetitive topics. Subsets with minimum cluster size 1 and cosine distance \(0.05-0.1\) show increased perplexity because the most credible topics are reduced to a small number of clusters in comparison to the topics that have not been clustered. Since a symmetric prior is used to compute perplexity, the uncertain topics outweight the credible topics. More interestingly, subsets of minimum clusters size 5, 10, or 20 show significantly better perplexity, depending on the cosine distance threshold, for instance, the subset of cluster topics with a minimum cluster size of 5 and at cosine distance larger than 0.15. The coherence plot (top right) and distinctiveness plot (bottom left) shows that highly recurrent topics (with minimum cluster size 10) tend to be more coherent and distinctive. We also observe that measures of coherence and distinctiveness decrease when including topics of lower recurrence or when increasing the cosine distance (letting more clusters be merged, so the new cluster grows in size). Interestingly, the credibility plot (bottom right) shows that the most credible subsets are formed with clusters of size 10 or more. Subsets of a minimum cluster size of 20 or cosine distance \(\leq 0.1\) are formed by a reduced number of clustered topics as shown in Figure \ref{ClusterNumber}. These topics may not repeat with the same certainty in other samples, and therefore, subsets with a small number of clusters tend to show high variability. Similar patterns are found when clustering LDA samples with 100 and 200 topics as shown in Appendix \ref{Clustering_LDA}.
  
Figure \ref{ClusterNumber} shows the number of clustered topics obtained by varying cosine distance thresholds and minimum cluster size. Subsets with a minimum cluster size of 1 report a large number of clusters (more than 100), which for visualization purposes are not shown. Topics that reappear in 20 samples are always fewer than 100 (number of topics of the LDA samples), confirming the uncertainty and low credibility of some topics.

Based on this analysis, we select a subset generated by minimum cluster size 10 and 0.35 CD threshold. Minimum cluster size 20 may lead to greater coherence but lower perplexity and vice versa minimum cluster size 1 or 5 leads to better perplexity but worse coherence. After the 0.35 CD threshold, perplexity is no longer improved. Both thresholds are also used to select a subset of clustered topics obtained from 100-topic LDA samples, and 0.45 CD for clustered topics obtained from 200-topic LDA samples. 

We repeat the 3 experiments with LDA samples with 50, 100 and 200 topics, but this time, we allow merging of topics within the same posterior sample. This implies that the clustering is no longer just a summary of the posterior distribution, but it is also, in effect, informing the number of topics within LDA. This allows us to compare and interpret some of the behaviour of the clustered topics from models with a large number of topics, as gathering similar topics from the same and different samples will form more distinctive subsets of clustered topics.

\begin{table}[h]
% is used to refer this table in the text
\centering
 \caption{Generalization, coherence, distinctiveness and stability metrics of LDA samples and subsets of clustered topics (HC-LDA and HC-LDA-WS) obtained from clustering LDA samples with 50, 100 and 200 topics.\label{table:BagComparison}}
\resizebox{\columnwidth}{!}{%
\renewcommand{\arraystretch}{2}
\small
 \begin{tabular}{|c c c c c c|} 
 \hline
   \multirow{1}{*}{{ }} 
  & \multirow{1}{*}{{ }} 
  & \multicolumn{1}{c}{{Generalization}} 
  & \multicolumn{1}{c}{Coherence} 
  & \multicolumn{1}{c}{Distinctiveness} 
  & \multicolumn{1}{c|}{Credibility} \\
  \cline{3-6}
   Model & Topics &  Perplexity & NPMI  & CD\(_{min}\) & \( \overline{\textrm{CS}_{\max}}\) \\
    & & Mean $\pm$ SE & Mean $\pm$ SE & Mean \(_{min}\) $\pm$ SE& Mean $\pm$ SE\\
     \hline
 LDA-50& 50 & 8.130 $\pm$ 0.003 & 0.325 $\pm$ 0.006 & 0.672 $\pm$ 0.020 & 0.769 $\pm$ 0.011\\ % inserting \
 
HC-LDA-50 & 52& \textbf{8.079} $\pm$ 0.006 & \textbf{0.333} $\pm$ 0.006 & 0.580 $\pm$  0.023 &    \textbf{0.916}  $\pm$ 0.014\\ %

HC-LDA-WS-50 & 50& \textbf{8.083} $\pm$ 0.005 & \textbf{0.333} $\pm$ 0.006 & 0.601 $\pm$  0.021 &    \textbf{0.907}  $\pm$ 0.014\\ %
 \hline
 LDA-100& 100 & 8.131 $\pm$ 0.003 & 0.319 $\pm$ 0.006 & 0.674 $\pm$ 0.016 & 0.716 $\pm$ 0.009\\ % inserting \
 
HC-LDA-100 & 96& \textbf{8.076} $\pm$ 0.006 & \textbf{0.333} $\pm$ 0.005 & 0.565 $\pm$  0.021 &    \textbf{0.890}  $\pm$ 0.010\\ %

HC-LDA-WS-100 & 86& \textbf{8.086} $\pm$ 0.005 & \textbf{0.331} $\pm$ 0.005 & 0.621 $\pm$  0.018 &    \textbf{0.882}  $\pm$ 0.012\\ %
 \hline
 LDA-200& 200 & 8.145 $\pm$ 0.003 & 0.302 $\pm$ 0.004 & 0.688 $\pm$ 0.011 & 0.644 $\pm$ 0.008\\ % inserting \
 
HC-LDA-200 & 198& \textbf{8.078} $\pm$ 0.005 & \textbf{0.32} $\pm$ 0.004 & 0.555 $\pm$  0.014 &    \textbf{0.864}  $\pm$ 0.007\\ %

HC-LDA-WS-200 & 145& \textbf{8.132} $\pm$ 0.003 & \textbf{0.335} $\pm$ 0.005 & 0.664 $\pm$  0.011 &    \textbf{0.848}  $\pm$ 0.011\\ %
 \hline
 \end{tabular}

}
\end{table}

In Table \ref{table:BagComparison}, we compare the performance of the selected subsets when topics from different samples form a cluster (HC-LDA), and when topics from the same and different samples form a cluster (HC-LDA-WS), against the average performance of the LDA models. As observed, subsets of clustered topics show significantly lower measures of generalization, larger topic coherence and larger topic credibility than LDA inferred topics. Note that topic distinctiveness is not improved, which might result from excluding highly distinctive non-recurrent topics. Allowing the merging of topics from the same samples retrieves fewer topics, does not significantly improve perplexity but increases the subset distinctiveness.

\begin{figure}[H]
    \centering
     \begin{subfigure}[b]{0.45\textwidth}
     \caption{HC-LDA-50 vs - HC-LDA-100}
         \centering
     \includegraphics[width=1\textwidth]{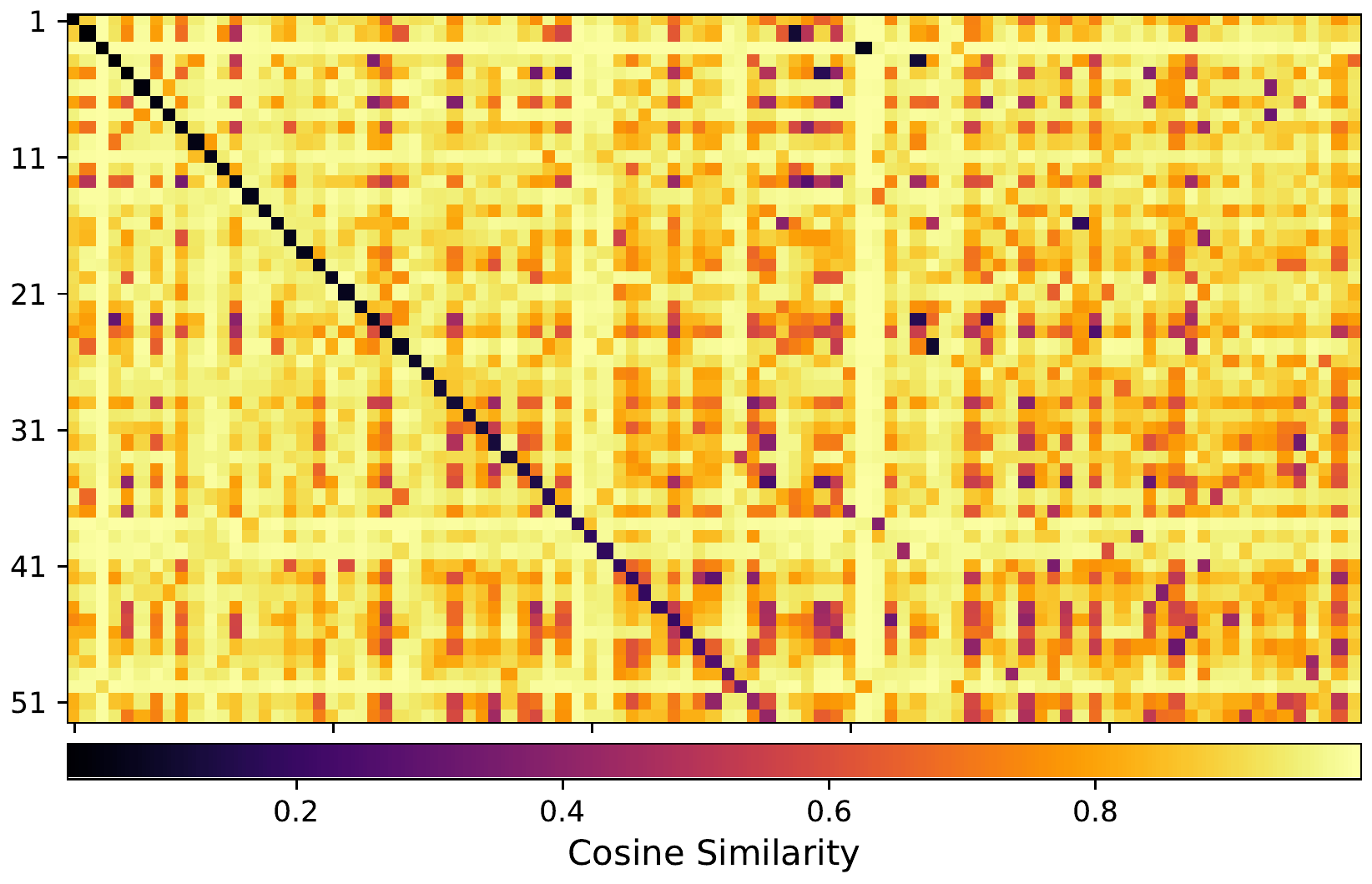}
     \end{subfigure}%
      \begin{subfigure}[b]{0.54\textwidth}
    \caption{HC-LDA-100 vs - HC-LDA-200}
        \centering
     \includegraphics[width=0.94\textwidth]{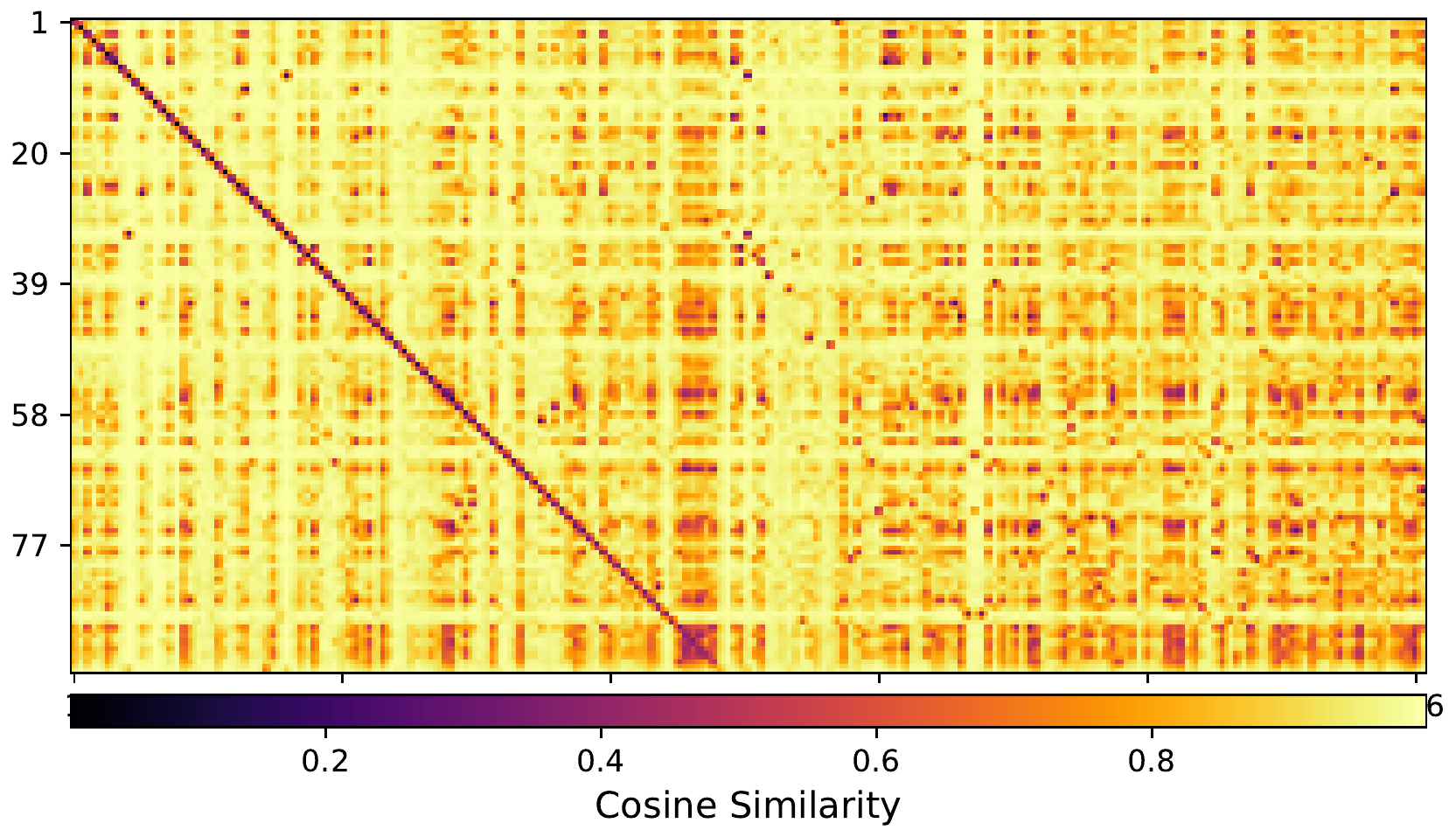}
     \end{subfigure}%
         \caption{Clustered topics correspondence between clustering of LDA samples with 50, 100 and 200 topics.}
        \label{correspondence}   
  \end{figure}

Different numbers of topics may retrieve similar performance. For example, Table \ref{table:BagComparison} shows that the subsets of clustered topics achieve similar average measures of perplexity, coherence and credibility; LDA models with 50 and 100 topics show the same levels of perplexity, coherence and distinctiveness. However, LDA samples with a large number of topics (and thereby their derived clustered topics) cover a wider variety of topics, highlighting important customer behaviours. For example, the Scottish topic illustrated in Figure \ref{t_8} is only found in LDA samples with 200 topics. Besides, clustered topics may be included in a subset derived from larger LDA samples. For instance, Figure \ref{correspondence} shows that the clustered topics in HC-LDA-50 (obtained from clustering 50-topic LDA samples) are also identified among the clustered topics in HC-LDA-100 (derived from 100-topic LDA samples). The latter is also identified among the clustered topics in HC-LDA-200 (derived from 200-topic LDA samples). Thus, the analysis of clustered topics obtained from LDA topics with a large number of topics may be warranted if the results reveal topics of interest, and the application of our clustering methodology can alleviate poor generalization for the over-parameterized model.

\section{Topics in British Grocery Retail\label{sec:grocerydiscussion}}

\begin{figure}
  \centering
  \begin{subfigure}[b]{0.32\textwidth}
    \centering
    \caption{Organic Food}
    \includegraphics[width=1\textwidth]{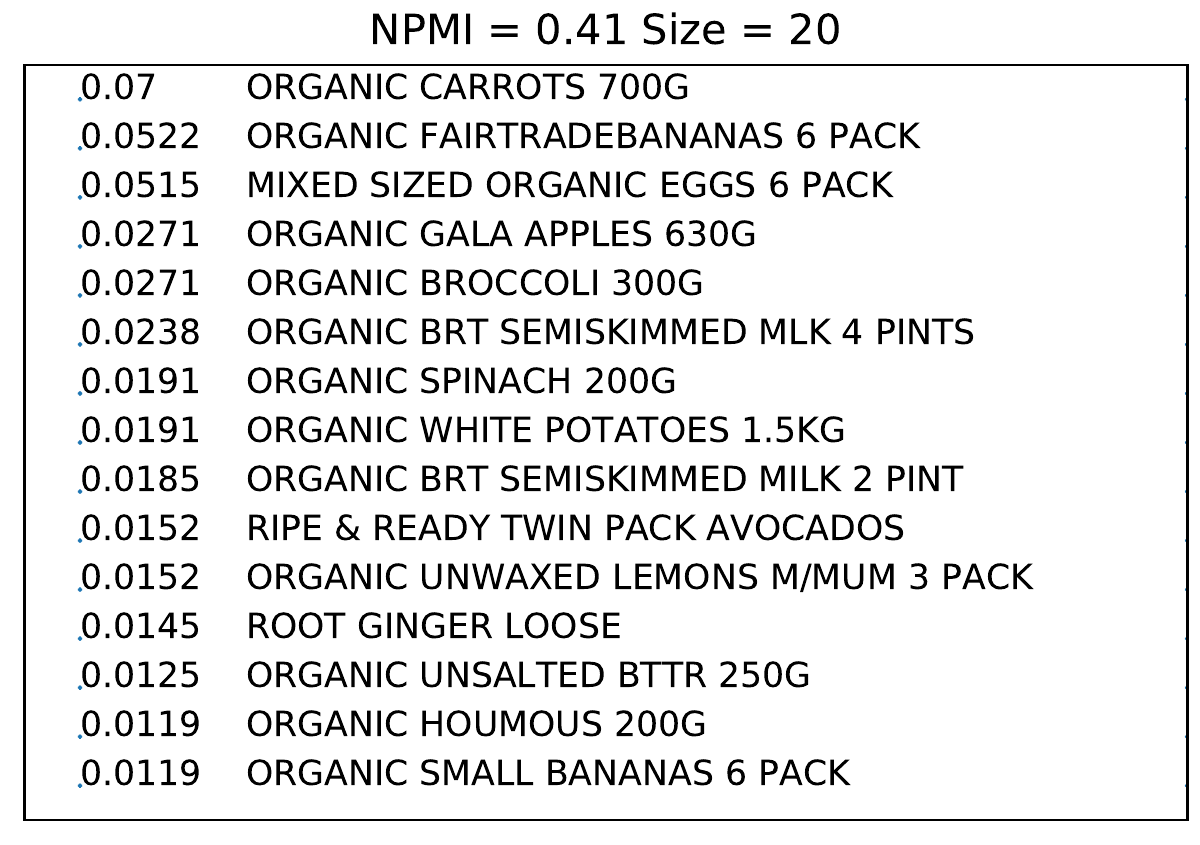}
    \label{t_1}
  \end{subfigure}
  \begin{subfigure}[b]{0.32\textwidth}
    \centering
    \caption{Italian dish}
    \includegraphics[width=1\textwidth]{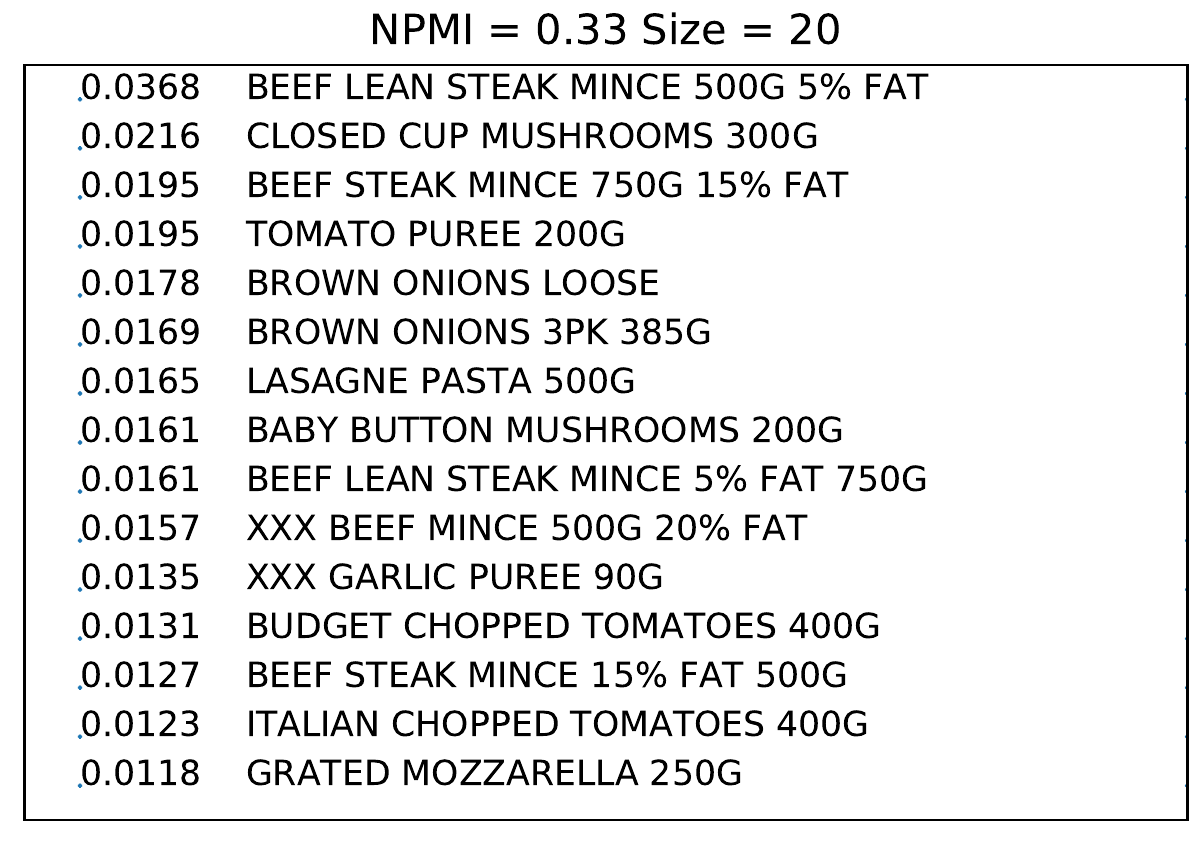}
    \label{t_2}
  \end{subfigure}
  \begin{subfigure}[b]{0.32\textwidth}
    \centering
    \caption{Gin and Tonic}
    \includegraphics[width=1\textwidth]{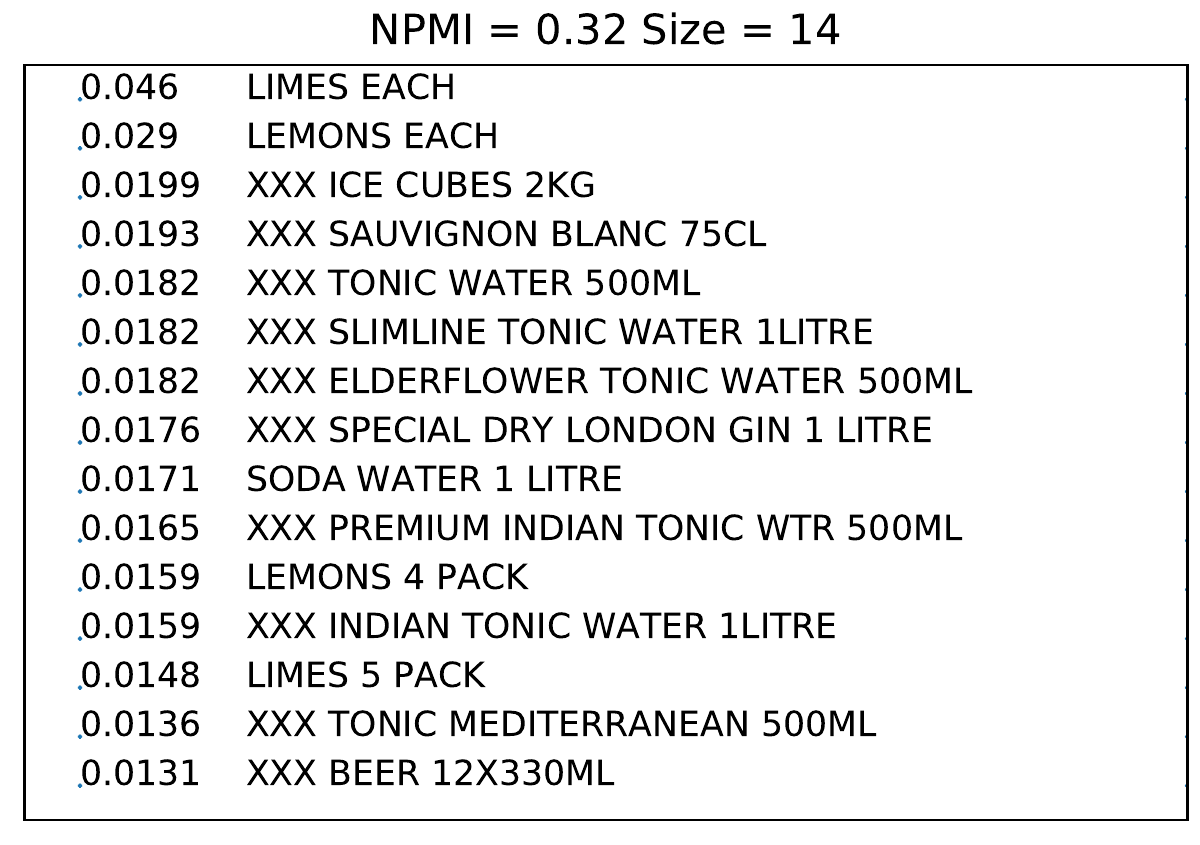}
    \label{t_3}
  \end{subfigure}
  \\
     \begin{subfigure}[b]{0.32\textwidth}
    \centering
    \caption{Meal promotion}
    \includegraphics[width=1\textwidth]{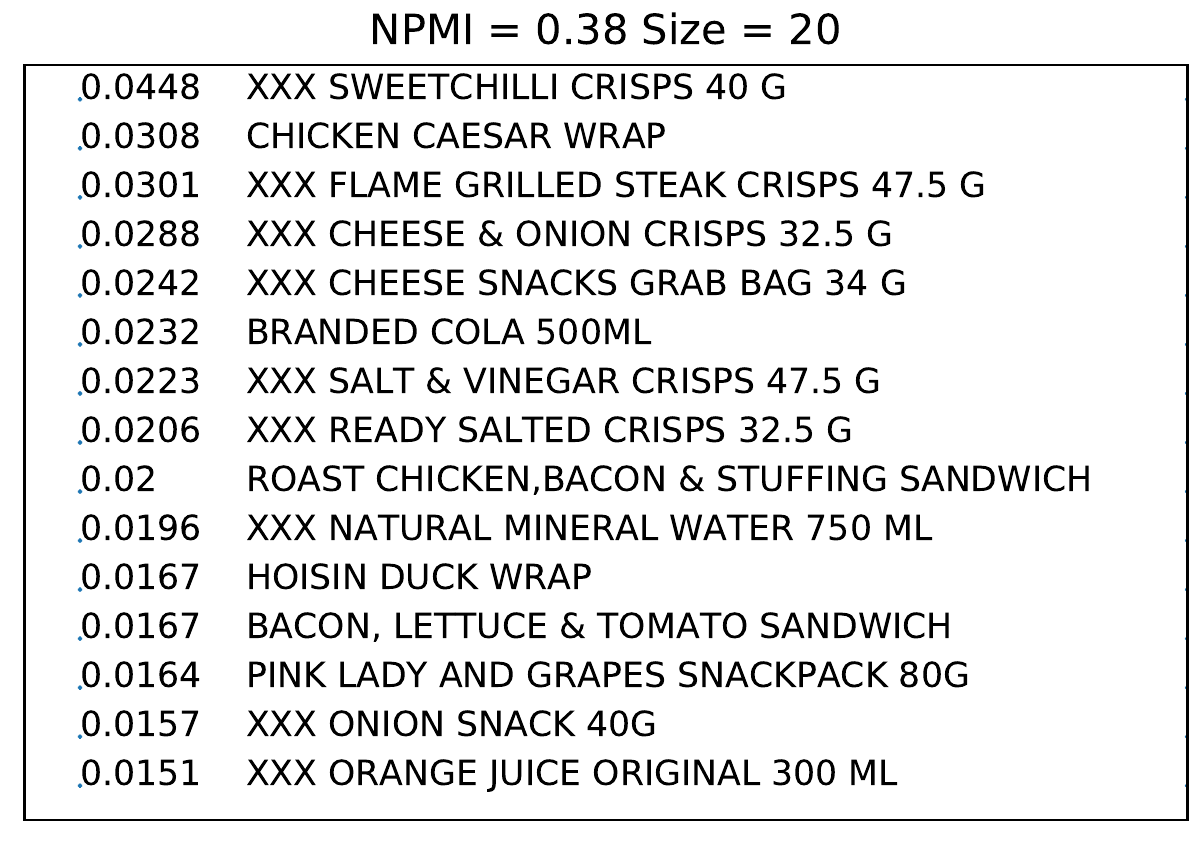}
    \label{t_4}
  \end{subfigure}
  \begin{subfigure}[b]{0.32\textwidth}
    \centering
    \caption{Budget line}
    \includegraphics[width=1\textwidth]{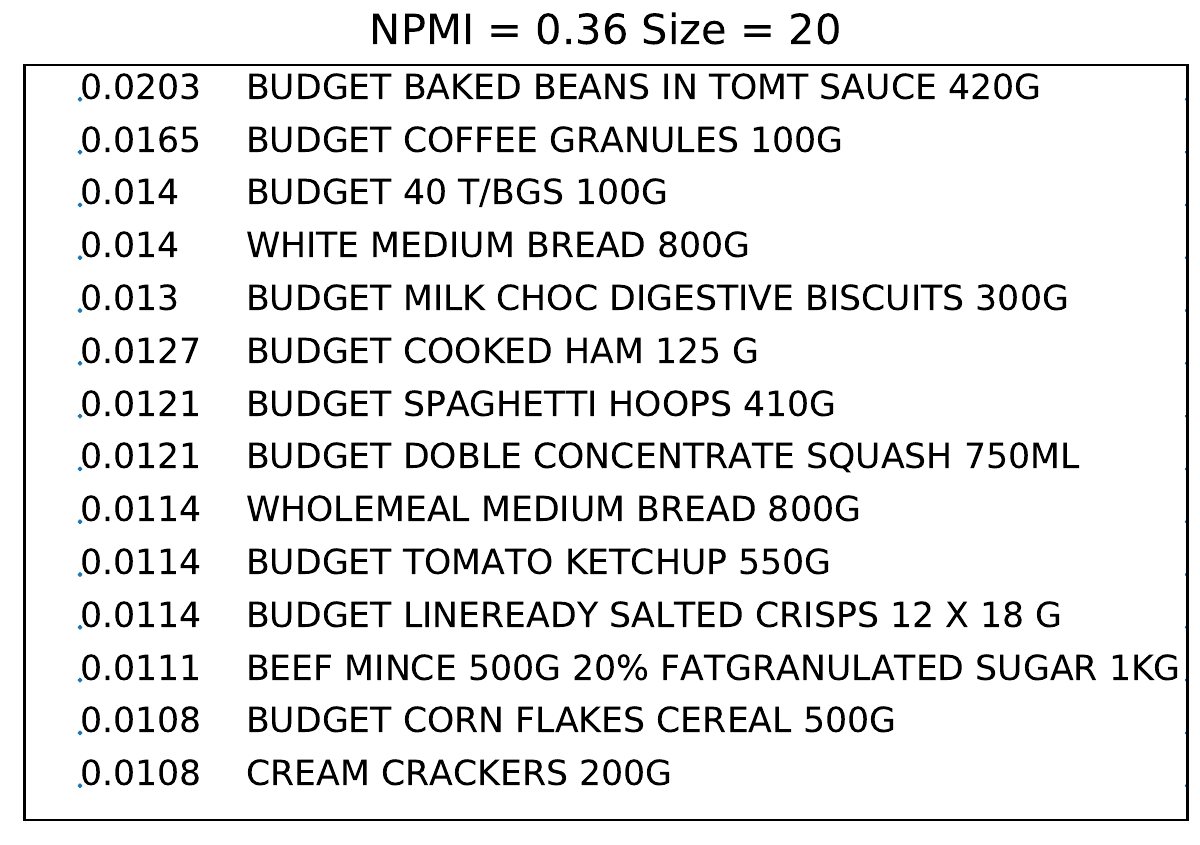}
    \label{t_5}
      \end{subfigure}
  \begin{subfigure}[b]{0.32\textwidth}
    \centering
    \caption{Dog goods}
    \includegraphics[width=1\textwidth]{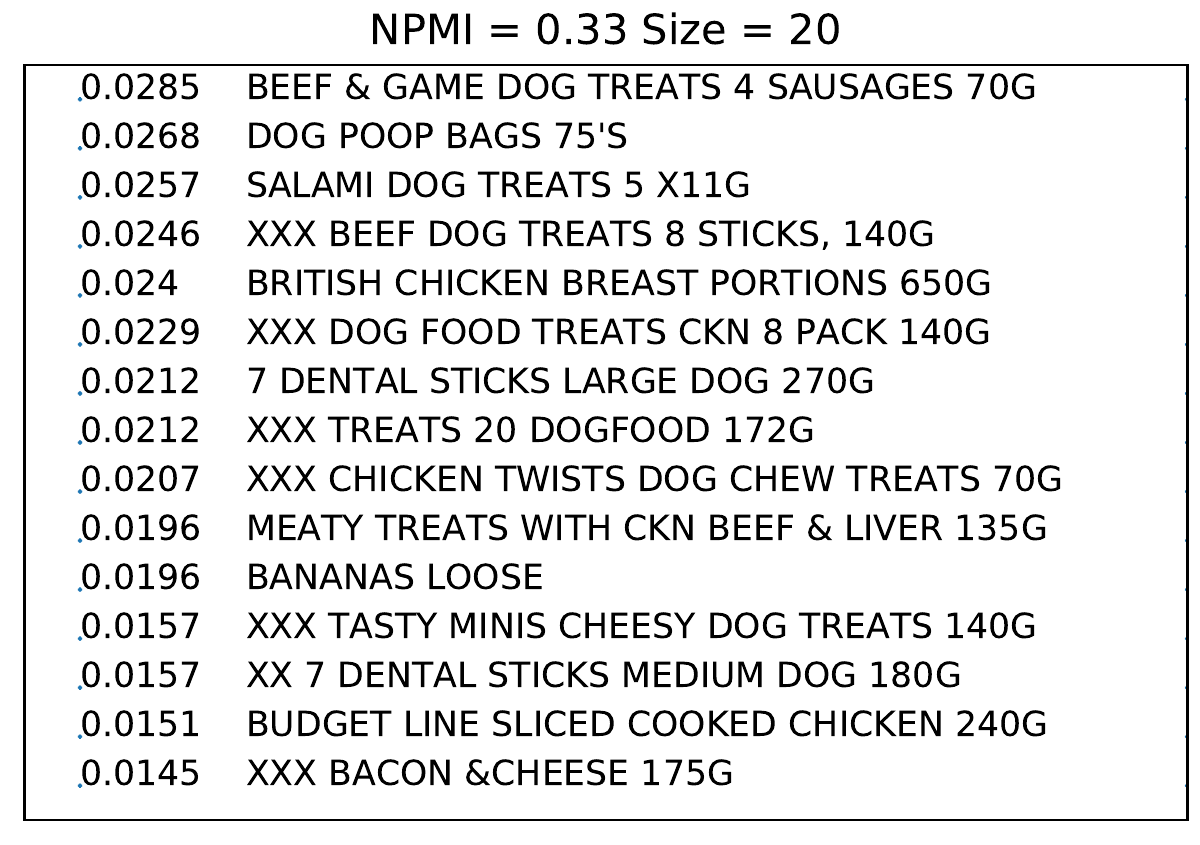}
    \label{t_6}
  \end{subfigure}
  \\
  \begin{subfigure}[b]{0.32\textwidth}
    \centering
    \caption{Roast Dinner}
    \includegraphics[width=1\textwidth]{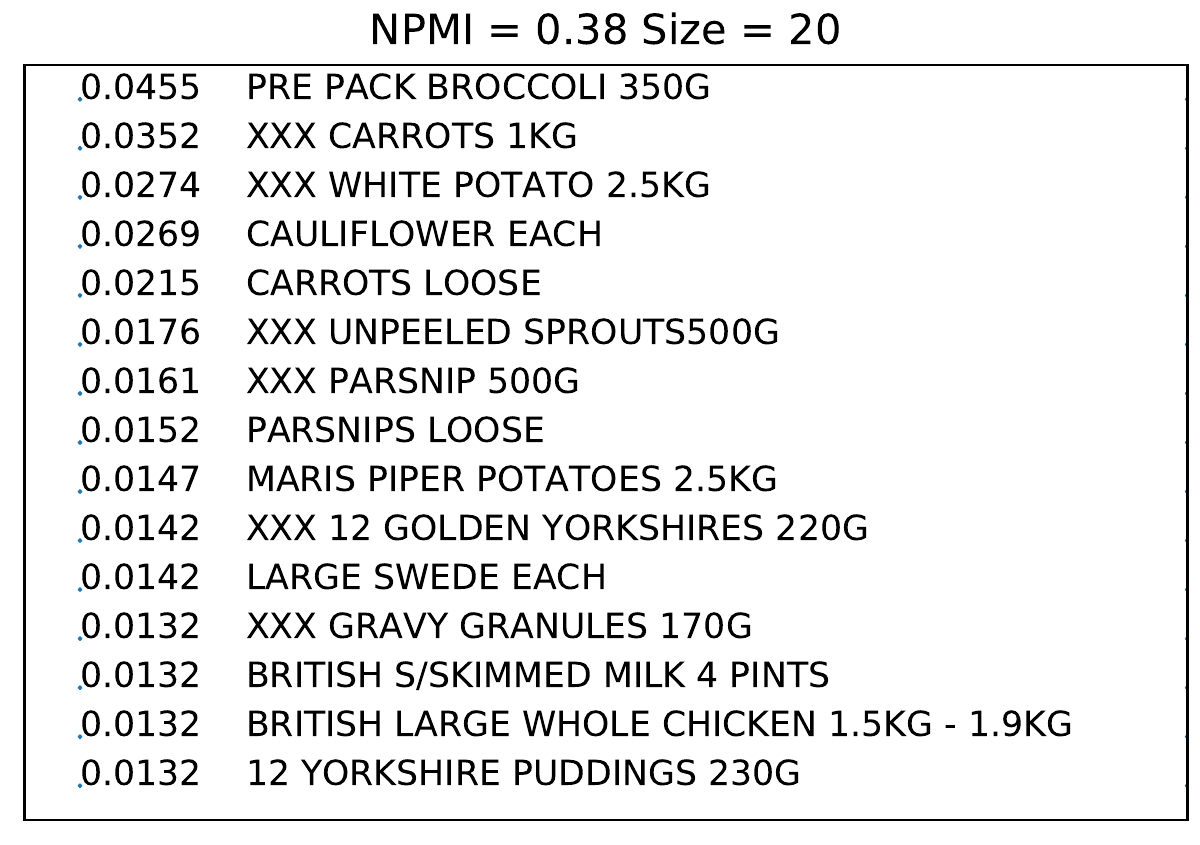}
    \label{t_7}
  \end{subfigure}
    \begin{subfigure}[b]{0.32\textwidth}
    \centering
    \caption{Scottish topic}
    \includegraphics[width=1\textwidth]{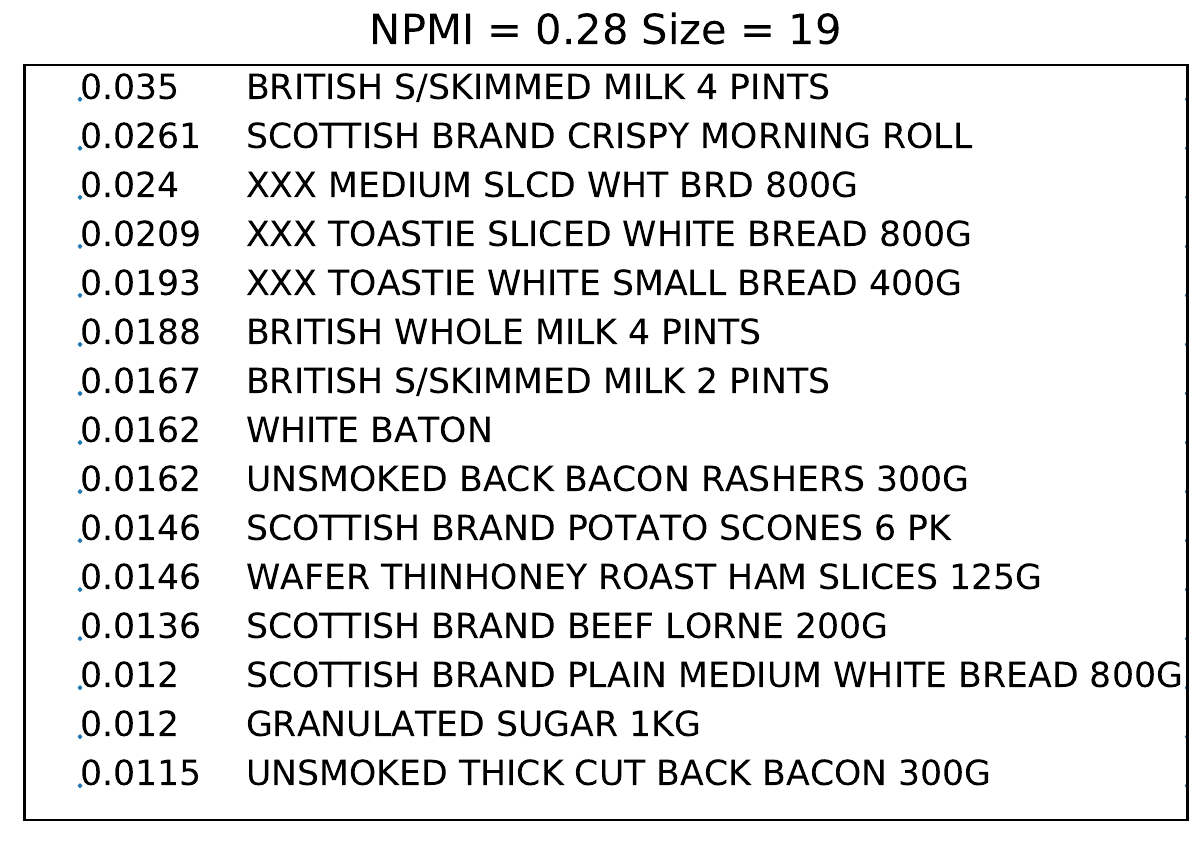}
    \label{t_8}
  \end{subfigure}
    \begin{subfigure}[b]{0.32\textwidth}
    \centering
    \caption{Christmas essentials}
    \includegraphics[width=1\textwidth]{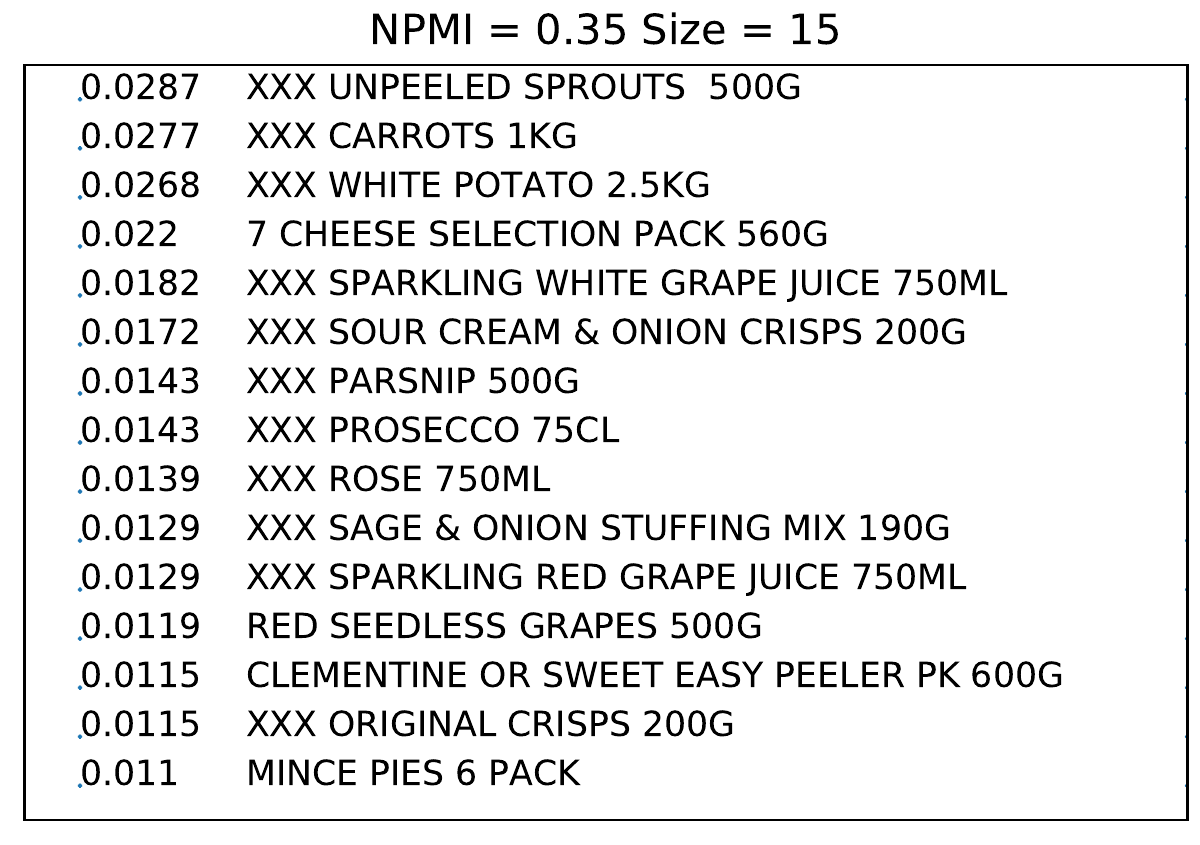}
    \label{t_9}
  \end{subfigure}            
  \caption{Topics in the UK grocery retail market baskets. Each topic is characterized by the 15 products with the largest probabilities. Probabilities and products are sorted in descending order. Brand names have been replaced by XXX for anonymization purposes. Credibility shows the ratio between the number of topics distributions associated with each cluster and the number of posterior draws. Topics reflect a variety of shopping motivations, i.e., diet orientations, cooking from scratch, ready-to-eat meals, preference for budget/premium product lines, pet ownership/household composition, specific events, geography and seasonality. Topics may also be associated with alcohol/fat/salt/sugar consumption.}
  \label{topics_uk}
\end{figure}

The analysis of topics and the products that together fulfil customers' motivations convey customer insights, i.e., diet orientations, cooking from scratch, preference for specific drinks or dishes, etc. For instance, Figure \ref{t_1} presents the topic of `Organic Food', Figure \ref{t_2} shows ingredients to cook an `Italian dish' and Figure \ref{t_3} highlights ingredients to prepare `Gin and Tonic'. Along with these topics, other identified topics show vegetarian-friendly foods, free-from lactose/gluten foods, ingredients for cooking Asian, Mexican, or Indian recipes. In these examples, topics gather products from different categories, i.e, ice and tonic water are two categories while tonic water and soda water are in the same category. Identifying combinations of products from different categories may have useful and commercial implications in improving product recommendations, developing promotional campaigns, optimizing assortments and planning shelf space, etc.

In contrast to cooking from scratch, customers may prefer convenience foods such as ready-to-eat meal promotions. For example, Figure \ref{t_4} represents a `meal promotion' composed of a sandwich, a bottle of soda or water, and a package of prepared fruit or crisps. Topics also show that customers tend to choose products within the supermarket's budget line or premium line, e.g., Figure \ref{t_5} gathers products from a `budget line' which offers products of a lower price than branded substitutes. Pet-ownership or household composition can be illustrated by topics, for instance, Figure \ref{t_6} lists `dog goods', including food, meat, and cleaning items. Other topics illustrate baby-related foods and large size items indicating household composition. Topics reveal customer's decision drivers, which can aid further customer analysis such as customer segmentation and customer profiling, to improve customer experience and to build brand loyalty.

Topics reveal customer motivations that are driven by specific events, geography, or seasonality. For instance, Figure \ref{t_7} depicts the `roast dinner' which is a traditional British main meal that is typically served on Sunday. Other event-specific topics manifest customers' motivations such, as having a picnic, buying a gift (flowers and chocolates), or barbecue. Topics also exhibit specific shopping themes that are driven by products that are available or highly preferred in certain locations or at specific periods. For example, Figure \ref{t_8} reveals Scottish-branded products in the `Scottish topic'. Similarly, a Northern Irish topic includes packed and locally supplied foods. Figure \ref{t_9} shows the `Christmas essentials' topic which is characterized by mince pies, sparkling grape juice, produce, and snacks. Easter and Halloween are also depicted by topics that contain the icons: chocolate egg and pumpkin, respectively. Commercially speaking, identifying events and geographical/seasonal patterns may inform marketing campaigns and support the optimization of product assortment.

Our approach allows us to provide measures of uncertainty for each inferred topic. For example, the topics `Organic food', `Italian dish' appeared in every single posterior draw. Therefore, corresponding commercial decisions can be made with relative confidence in these shopping themes. On the other hand, less frequent topics can be identified. For instance, the topics `Scottish' and `Christmas essentials' appeared 19 out of 20 and 15 out of 20 times, respectively, within the 20 LDA posterior draws. The lower frequency of these topics might be explained by the small representation of them in our data due to their regional/seasonal nature. More importantly, naive averaging of posterior draws would have damaged these topics by merging them with an irrelevant topic. 

Understanding grocery consumption not only assists marketing practices but also opens up new avenues for social research. Uncovering consumption patterns related to alcohol (Figure \ref{t_3})/fat/sugar/salt through topic modelling is scalable, low-cost and allows the identification of specific products and their characteristics. Thus, topic modelling may help the conduction of dietary studies that are typically limited to survey data such as food frequency questionnaires and open-ended dietary assessment \citep{aiello2019large,Einsele2015ASA,wang2014fruit,wardle2007eating}.

\section{Conclusion\label{sec:conclusion}}

In this paper, we expand the evaluation process of LDA to include qualitative aspects such as topic coherence, topic distinctiveness, and topic credibility along with model generalization. In addition, we propose a methodology that post-processes LDA models, to summarize the entire posterior distribution of an LDA model into a single set of topical modes. Our approach identifies recurrent topics using meaningful distance criteria and allows the user to assess topic credibility. The distance criteria were developed through a customized survey which we carried out with experts in the field of grocery retailing; these helped us evaluate and set thresholds that assist the evaluation of interpretability and similarity of grocery retail topics. Empirically, we showed the advantages of the proposed methodology in terms of capturing topic uncertainty and enhancing coherence and credibility. We identified credible and coherent topics that exhibit a variety of shopping motivations, i.e., diet orientations, cooking from scratch, specific events, pet ownership, geography, seasonality, etc. Topics can be associated with alcohol/fat/salt/sugar consumption, which may provide new venues for sociological research. Finally, our methods focused on the context of LDA models. Summarizing multiple posterior draws from a mixture model, however, is a challenge that extends beyond LDA. Our methods can be implemented beyond LDA by replacing the cosine distance with other measures relevant to each context.

% Appendix here Options are (1) APPENDIX (with or without general title) or (2) APPENDICES (if it has more than one unrelated sections) Outcomment the appropriate case if necessary
%
% \begin{APPENDIX}{<Title of the Appendix>}
% \end{APPENDIX}
%
% or
%
\clearpage
\begin{appendix}
  
\section{MCMC convergence}

\begin{figure}[H]
  \begin{subfigure}[b]{0.45\textwidth}
    \centering
    \caption{50 topics. \(\hat{R}: 1.00\) }
    \includegraphics[width=0.75\linewidth]{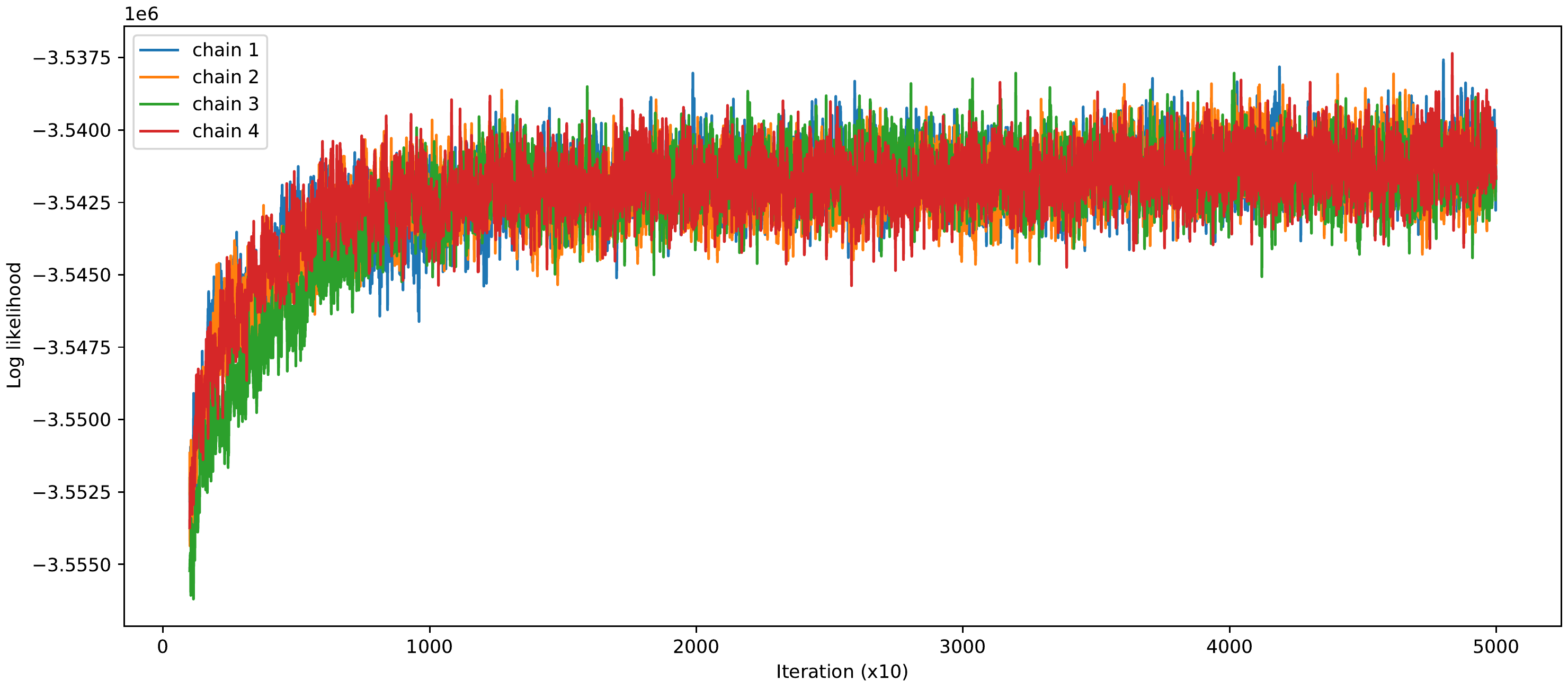}
    \label{conve_50}
  \end{subfigure}%
  \begin{subfigure}[b]{0.45\textwidth}
    \centering
    \caption{100 topics. \(\hat{R}: 1.10\) }
    \includegraphics[width=0.75\linewidth]{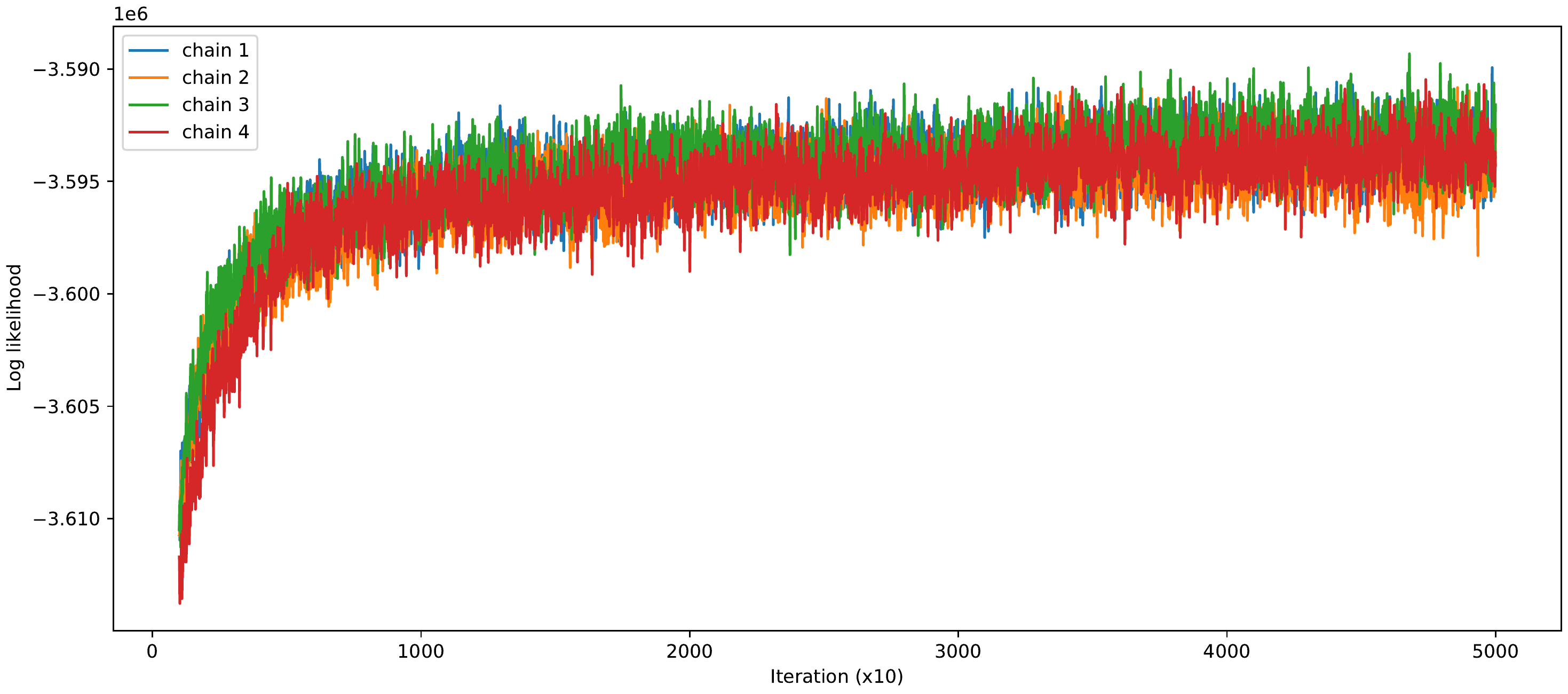}
    \label{conve_100}
  \end{subfigure}%
  \\
  \\
  \begin{subfigure}[b]{0.45\textwidth}
    \centering
    \caption{200 topics. \(\hat{R}: 1.02\) }
    \includegraphics[width=0.75\linewidth]{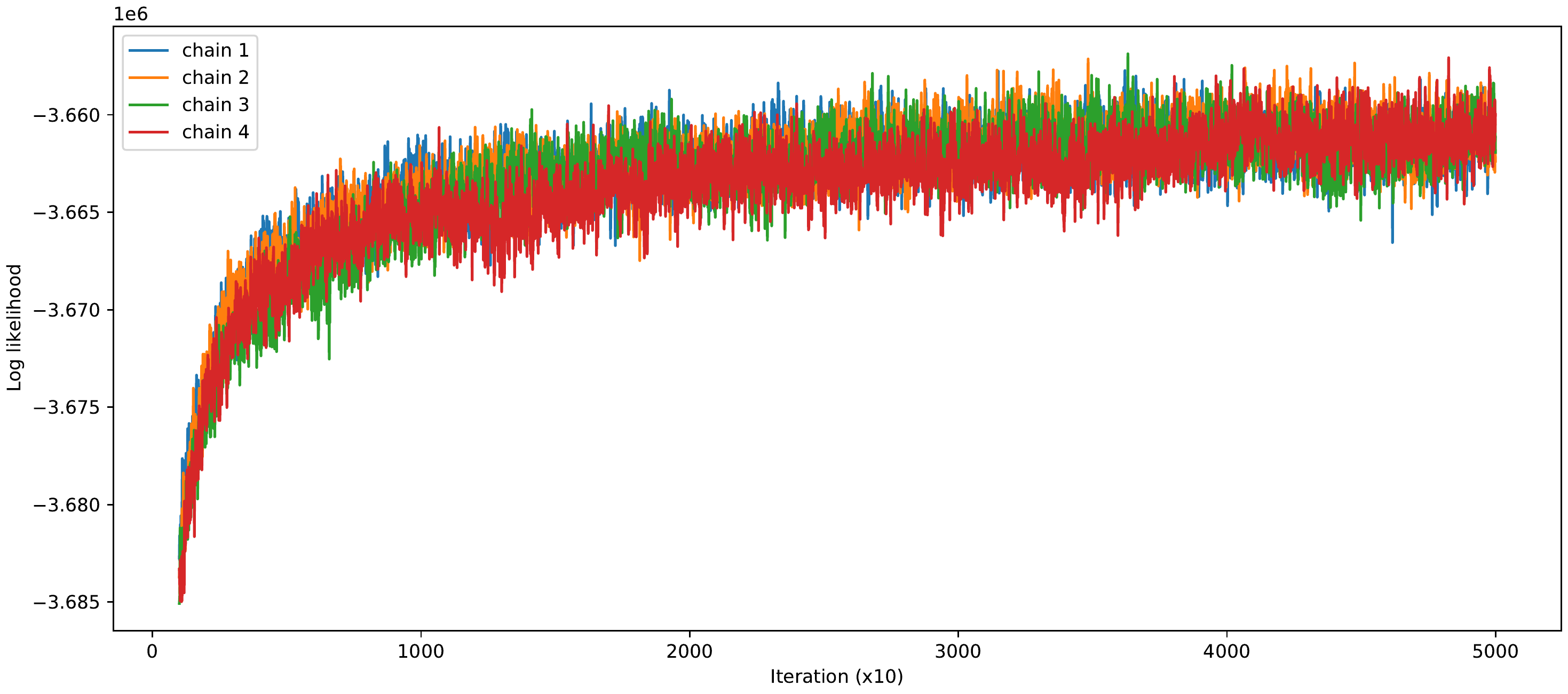}
    \label{conve_200}
  \end{subfigure}%
  \begin{subfigure}[b]{0.45\textwidth}
    \centering
    \caption{400 topics. \(\hat{R}: 1.13\) }
    \includegraphics[width=0.75\linewidth]{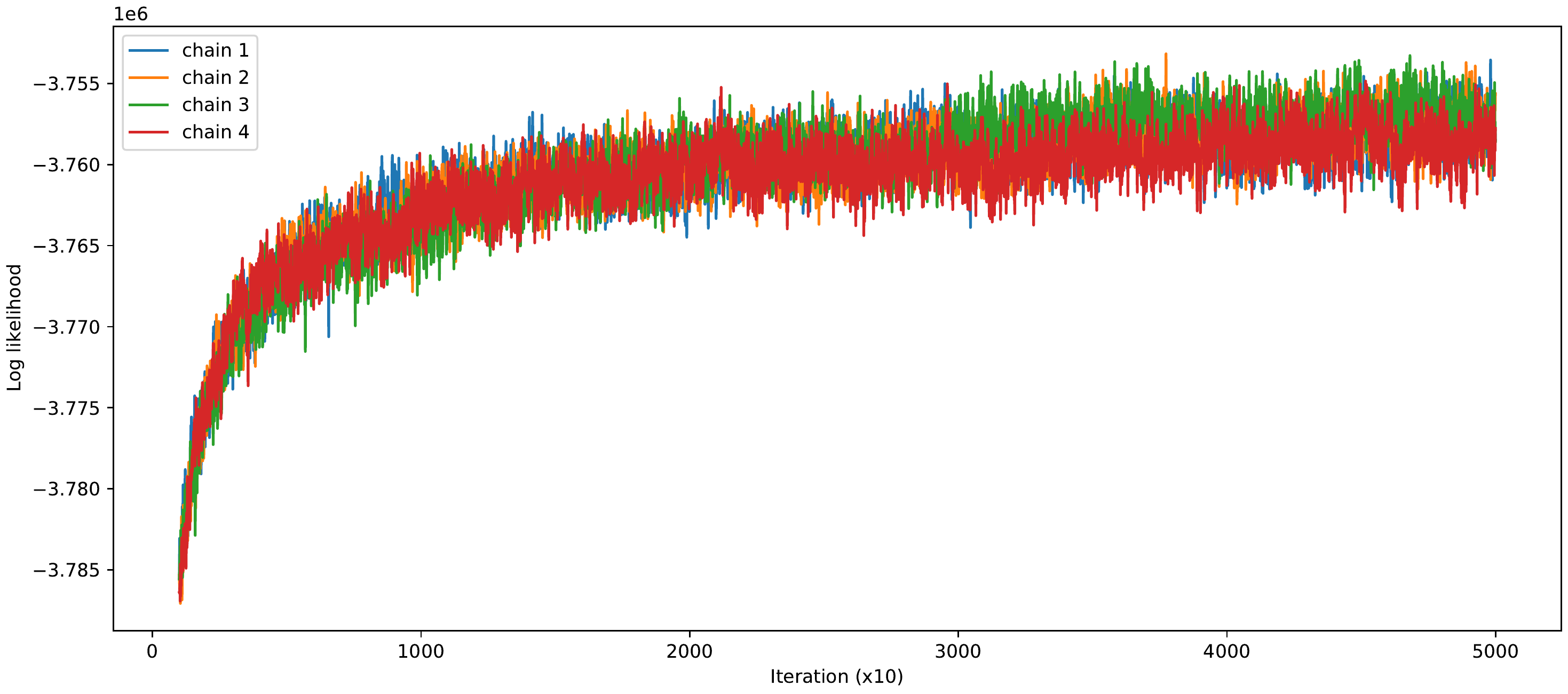}
    \label{conve_400}
  \end{subfigure}%
  \caption{Markov Chains of LDA with 50, 100, 200, and 400 topics. \(\hat{R}\) is the potential scale reduction factor.}
  \label{convergence}
    \end{figure}

For each LDA model, 4 Markov chains are run for 50,000 iterations with a burn-in period of 30,000 iterations. We evaluate convergence using the potential scale reduction factor \(\widehat{R}\) \citep{gelman2013bayesian}. When \(\widehat{R}\) is near 1, we can assume that samples approximate the posterior distribution. Values of \(\widehat{R}\) below 1.1 are acceptable. Figure \ref{convergence} shows the trace plot for the log-likelihood (measured at every 10 iterations) of LDA with 50, 100, 200 and 400 topics. We calculate the potential scale reduction factor using 4 chains and 8000 samples. Chains for
LDA with 50, 100, 200 topics seem to be converged. The chains for LDA with 400 topics need to be further trained, however, preliminary evaluation of the topics from these chains already show lower performance than topics from chains with fewer topics.
 
\label{mcmc_convergence}

\section{Clustering of topics}
\label{Clustering_LDA}

Here, we show the evaluation of subsets of clustered topics obtained from 20 LDA posterior samples with 100 and 200 topics.

\begin{figure}[H]
    \centering
    \begin{subfigure}[b]{0.45\textwidth}
      \caption{Clustering with 100-topic LDA chains}
     \includegraphics[width=1\textwidth]{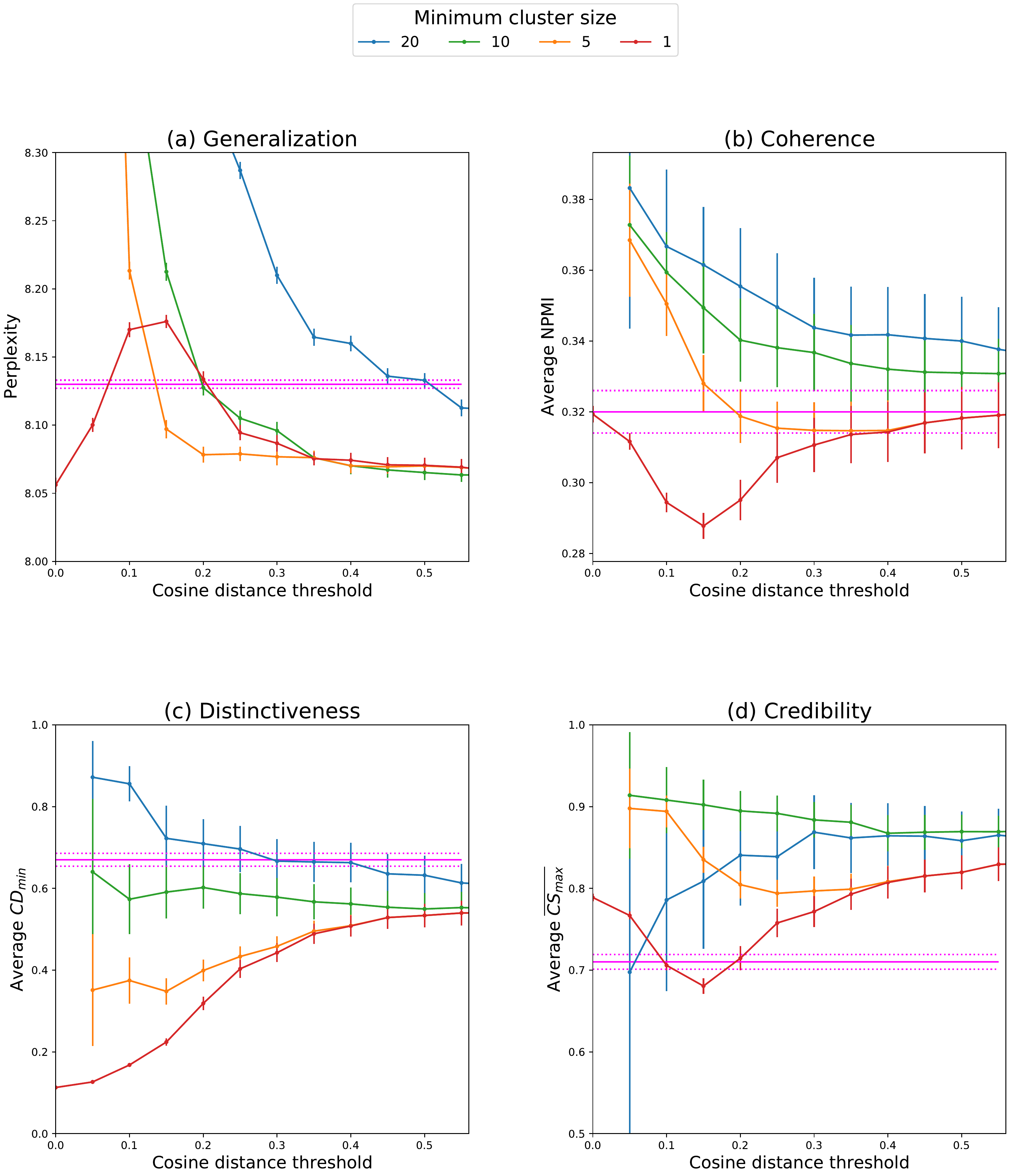}
     \end{subfigure}%
     \begin{subfigure}[b]{0.45\textwidth}
       \caption{Clustering with 200-topic LDA chains}
     \includegraphics[width=1\textwidth]{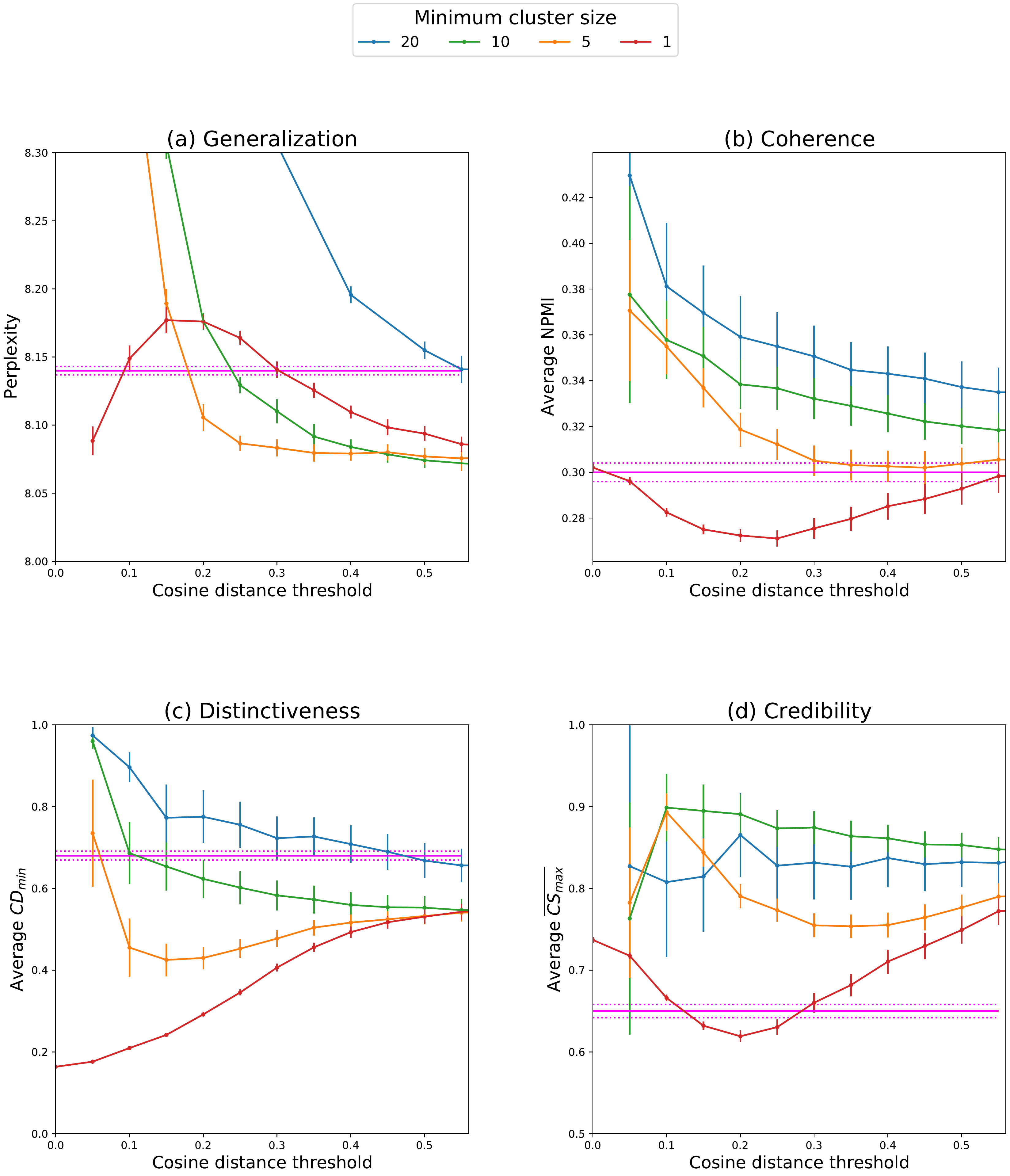}
     \end{subfigure}%
         \caption{Subset evaluation using cosine distance (varying from 0 to 1 with increments of 0.05) and minimum cluster size (20, 10, 5 and 1). Clustered topics were obtained from clustering 20 samples of LDA with 100/200 topics. Vertical lines represent one standard error. Magenta lines show the average measures (\(\pm\) one standard error) of the LDA samples.}
        \label{EvaluationClustering}
  \end{figure}

  \begin{figure}[H]
    \centering
    \begin{subfigure}[b]{0.45\textwidth}
    \caption{100-topic LDA chains}
     \includegraphics[width=1\textwidth]{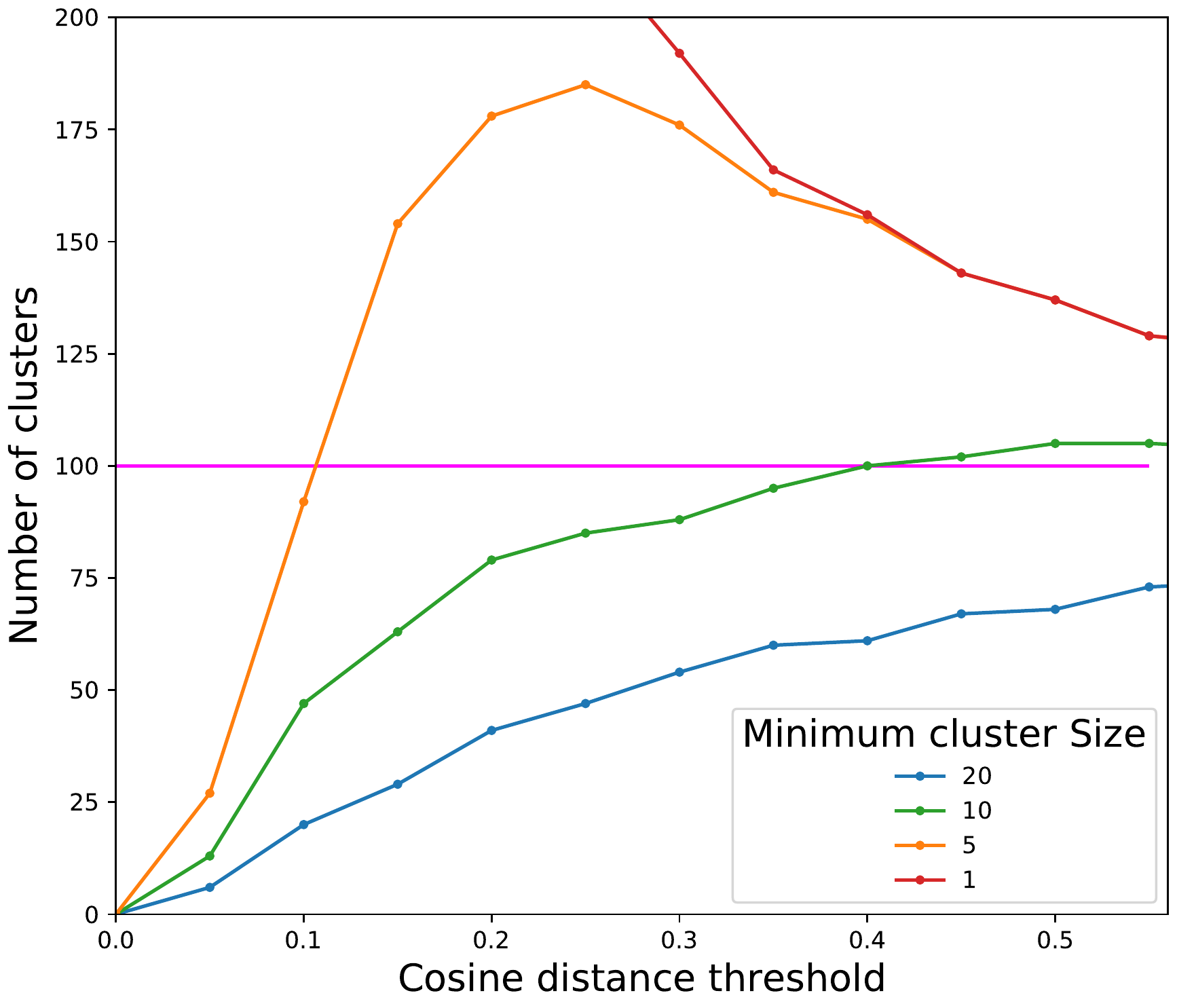}
     \end{subfigure}%
     \begin{subfigure}[b]{0.45\textwidth}
     \caption{200-topic LDA chains}
     \includegraphics[width=1\textwidth]{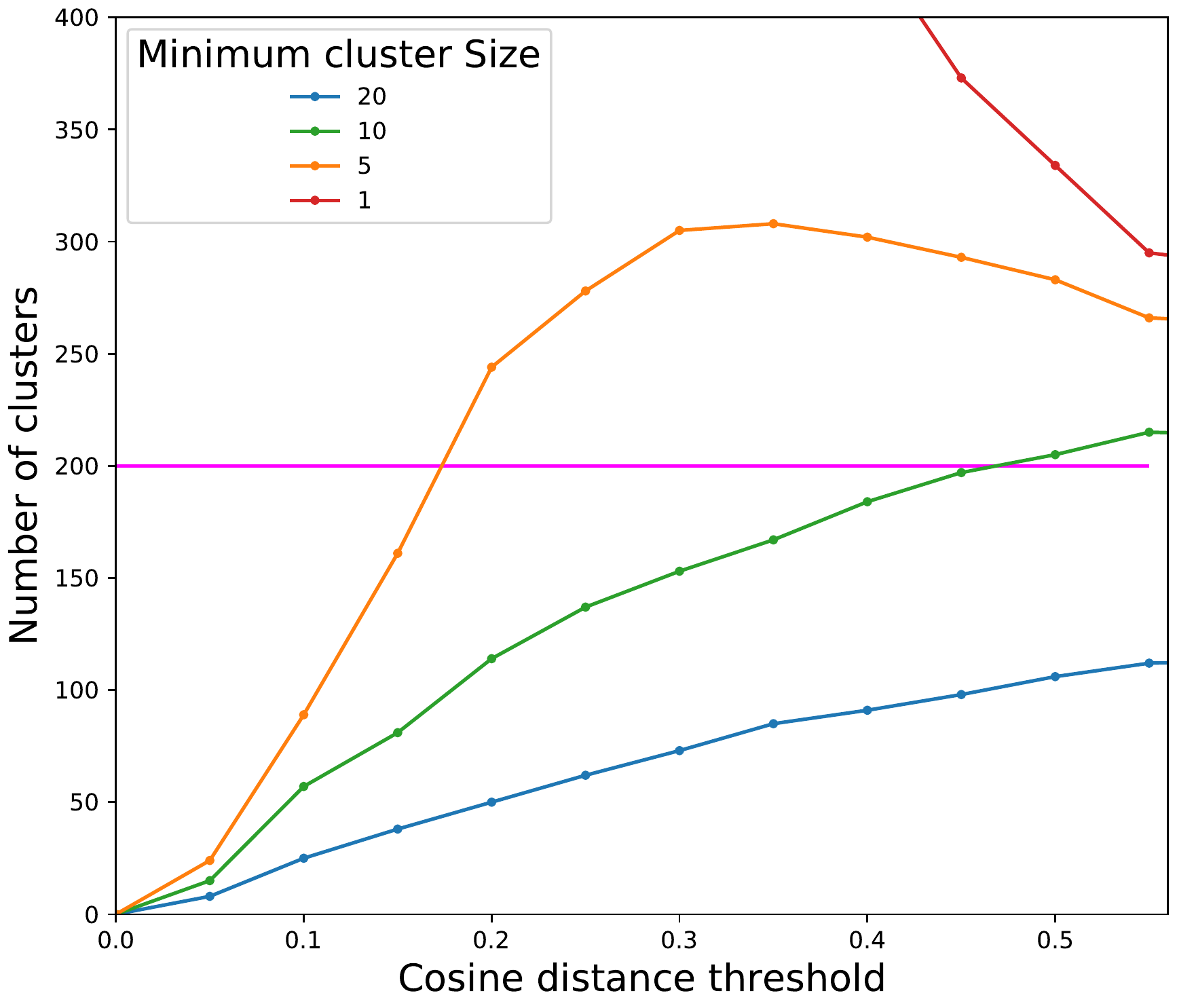}
     \end{subfigure}%
      \caption{The number of clusters obtained at cosine distance (varying from 0 to 1 with increments of 0.05) and minimum cluster size (20, 10, 5 and 1). The magenta line shows the number of topics in the LDA samples. For visualization purposes, subsets larger than 200/400 clusters are not shown.}
        \label{NClusteriong}
  \end{figure}

% etc
\end{appendix}

% Acknowledgments here \ACKNOWLEDGMENT{The authors gratefully acknowledge the existence of the Journal of Irreproducible Results and the support of the Society for the Preservation of Inane Research.}

% References here (outcomment the appropriate case)

% CASE 1: BiBTeX used to constantly update the references (while the paper is being written).  \bibliographystyle{informs2014} % outcomment this and next line in Case 1
% \bibliography{<your bib file(s)>} % if more than one, comma separated

% CASE 2: BiBTeX used to generate mypaper.bbl (to be further fine tuned) \input{mypaper.bbl} % outcomment this line in Case 2

% If you don't use BiBTex, you can manually itemize references as shown below.

\bibliographystyle{rss.bst} \bibliography{references.bib}

%%%%%%%%%%%%%%%%%
\end{document}